\documentclass[11pt]{article}
\usepackage[a4paper,margin=2.3cm]{geometry}
\usepackage[utf8]{inputenc}
\usepackage[T1]{fontenc}
\usepackage{amsmath,amssymb}
\usepackage{graphicx}
\usepackage{booktabs}
\usepackage{siunitx}
\usepackage{float}
\usepackage{algorithm}
\usepackage{algpseudocode}
\usepackage{xcolor}
\usepackage[colorlinks=true,linkcolor=blue!55!black,urlcolor=blue!55!black,citecolor=blue!55!black]{hyperref}
\usepackage[affil-it]{authblk}
\usepackage[nolist]{acronym}
\acrodef{MNC}{magnetic Newton's cradle}
\acrodef{NC}{Newton's cradle}
\acrodef{ODIL}{Optimizing a DIscrete Loss}
\acrodef{RMS}{root mean square}
\acrodef{FFT}{fast Fourier transform}
\acrodef{AD}{automatic differentiation}

\makeatletter\AtBeginDocument{\check@mathfonts}\makeatother
\newcommand{\Imag}{I_{\mathrm{mag}}}
\newcommand{\lcom}{l_{\mathrm{com}}}
\newcommand{\fg}[3][0.72\linewidth]{\begin{figure}[ht]\centering\includegraphics[width=#1]{figures/#2}\caption{#3}\end{figure}}

\title{\textbf{Data Driven Modeling of a Magnetic Newton's Cradle using the Optimization of  a Discrete Loss}\\[2pt]}

\author[1]{Dimitris Koumoutsakos
}
\author[2]{Sergey Litvinov}
\affil[1]{Lexington High School, 251 Waltham St, Lexington, MA 02421, United States}
\affil[2]{Computational Science and Engineering Laboratory, Harvard University, 23 Oxford St,
Cambridge, MA 02138, United States}
\date{}

\begin{document}
\maketitle
\vspace{-1em}

\begin{abstract}
We investigate the dynamics of a \ac{MNC} using experiments, simulations through forward and inverse data driven models. In the \ac{MNC}  magnetic forces replace the instantaneous elastic collisions of the classical design. We compare point-dipole and finite-current-loop interaction models, identifying their parameters from direct measurements as well as through  pendulum trajectories using the method of \ac{ODIL}. For a synthetic three-pendulum twin, \ac{ODIL} reproduces discretized trajectories to round-off accuracy and infers accurately magnet strength and damping from sparse data even when only one pendulum is observed. Model comparison shows that the point-dipole model compensates for near-field errors by underestimating the magnetic moment and produces structured residuals at each momentum transfer. The current-loop model removes this bias and, on full-precision recordings, also recovers the magnet radius. The continuous magnetic coupling reproduces the classical cradle's key behavior while enabling richer dynamics, including $N$ normal modes and a static repulsion offset.  These results demonstrate that \ac{ODIL} can infer physical parameters, and quantify model-form errors from sparse, noisy observations. Ultimately, this work establishes the magnetic Newton's cradle not merely as a physical curiosity, but as a rigorously quantifiable testbed for computational inverse modeling.
\end{abstract}
\acresetall

\tableofcontents
\clearpage

\section{Introduction}
\ac{NC} is a classical  demonstration of momentum and energy conservation in mechanical systems. In an \ac{NC}, a row of identical balls, hung by strings so that they touch at rest, transfers a released ball's momentum through the
chain via a sequence of elastic collisions. In the  ideal case, only the balls at the two ends are the ones moving. The \ac{MNC} replaces  the balls with permanent magnets mounted so that
neighbors \emph{repel} each other. In the \ac{MNC}, the momentum transfer by  collisions is replaced by a continuous magnetic force acting over a finite range. This continuous coupling in the \ac{MNC} reproduces the energy transfer of the classical \ac{NC} while introducing a richer dynamics. In the \ac{MNC}, we observe a full spectrum of normal modes and amplitude-dependent eigenfrequencies.\footnote{This 
system was posed as IYPT 2026 Problem 15 with the instruction to explain and study its motion.}

The behavior of the \ac{NC} and the \ac{MNC}  can be modeled  by a set of coupled differential equations. The key modeling question involves the formulation of the magnetic force between pendulums. The classical approach uses  point-dipole approximations. 
In~\cite{russell1997} the authors built an anharmonic
chain of magnetic pendulums as an experimental paradigm for \emph{moving breathers} described as 
spatially localized, dynamically stable oscillating states in a discrete nonlinear lattice
In~\cite{cuevas2009} the authors presented   a forced-damped array of
coupled pendulums and observed a stability reversal of intrinsic localized modes (discrete breathers) near a critical coupling.
Molerón et al.~\cite{moleron2014} studied a one dimensional chain of repelling magnets  and showed experimentally, numerically, and
theoretically that it supports amplitude-dependent solitary waves, with a smooth 
profile at low energy and single site localization at high energy. García-Raffi
et al.\ \cite{garciaraffi2018} studied nonlinear wave propagation with a similar chain of magnetically coupled pendulums,
reporting harmonic generation from the interplay of nonlinearity and dispersion. 

None of these works required  an accurate  near field force law, because
they were concerned with dynamics at moderate to large pendulum
spacing relative to magnet size, for which the dipole approximation is adequate. However, the point dipole approximation is not adequate at close spacings.
 Alternative formulations involve  modeling the magnets as finite current loops. On the magnetostatics side, closed-form expressions for the force between coaxial cylindrical
permanent magnets have been derived in terms of elliptic integrals~\cite{vokoun2009}, and 
via the equivalence between the amperian current model and a thin coil~\cite{ravaud2010}. Both results confirm that the force between
two axially-magnetized magnets scales as the square of the magnetization  at any separation, and reduces to the familiar $1/d^4$ dipole-dipole interaction law only at large separation.

In this paper, we investigate both models and their capability to reproduce the experimental data. We find that the two models differ already at the static gap of the cradle, where the pivot spacing is only a few magnet radii, and increasingly at the close approaches of larger swings. 

The models that we investigate involve also gravity terms as well as air drag. We identify all model parameters in two ways. First, by directly measuring the coefficients and then by extracting physical parameters such as the true magnetic field strength and microscopic rod asymmetries using inverse modeling through the pendulum trajectory data. Traditional inverse solvers that rely on forward numerical integration become fragile when the cradle reaches large amplitudes: the onset of chaos makes simulated trajectories diverge exponentially from the data for any small parameter error, and the  landscape seen by the optimizer becomes rugged\cite{baake1992fitting,ribeiro2020smoothness}. This paper treats the inverse dynamics of the magnetic Newton's cradle with the \ac{ODIL} framework~\cite{karnakov2024odil,karnakov2023flow}. The  discrete equations of motion are imposed as residuals of a least-squares problem whose unknowns are \emph{both} the system state at every time step and the hidden physical parameters. Because each residual couples only three neighboring time levels the Jacobian is sparse and Gauss--Newton steps are cheap. We use the framework in three ways. First, as an identification tool: on a synthetic twin of the three-pendulum cradle, with parameters representative of the apparatus of \S\ref{sec:apparatus}, we quantify how well the magnet strength and the damping can be recovered from sparse, noisy angle data (\S\ref{sec:odil-results}). Second, as a model-discrimination tool: fitting the point-dipole model to data generated by finite magnets exposes the model form error of the far field approximation as a bias in the identified moment and as a structured  residual, both of which vanish when the current-loop model is fitted instead. Third, as a calibration tool on the measured trajectories themselves: two independent recordings of the five-pendulum cradle and five of the three-pendulum cradle at spacings of 2--\SI{8}{cm} are fitted with both magnet models, with the initial state, per-rod natural frequencies and angle-calibration offsets treated as nuisance unknowns, and the identified parameters are compared with each other and with the direct measurements of the magnet strength (Hall probe, \S\ref{sec:hallprobe}), radius and damping (\S\ref{sec:odil-exp}). The same machinery is used to ask whether the amplitude-dependent damping of \S\ref{sec:damping} can be identified from the cradle's own motion. 

The paper is organized as follows. Section~\ref{sec:apparatus} describes the experimental apparatus, the two magnet models and the \ac{ODIL} methodology. Section~\ref{sec:linmodes} discusses the linear normal modes. Section~\ref{sec:results} presents the results: the direct measurements, the \ac{ODIL} identification and model discrimination on a synthetic twin (\S\ref{sec:odil-results}), the \ac{ODIL} calibration on the measured three- and five-pendulum trajectories, its repeatability and the comparison with the directly measured parameters (\S\ref{sec:odil-exp}), the closed-loop predictions for 2--5 pendulums, the mode spectra and the classical-cradle limit. The paper concludes with a discussion of the present findings, which together establish a calibrated digital twin of the magnetic Newton's cradle.


\section{Methods}
\label{sec:apparatus}

\subsection{Experimental apparatus}
\begin{figure}[H]\centering
\includegraphics[width=0.9\linewidth]{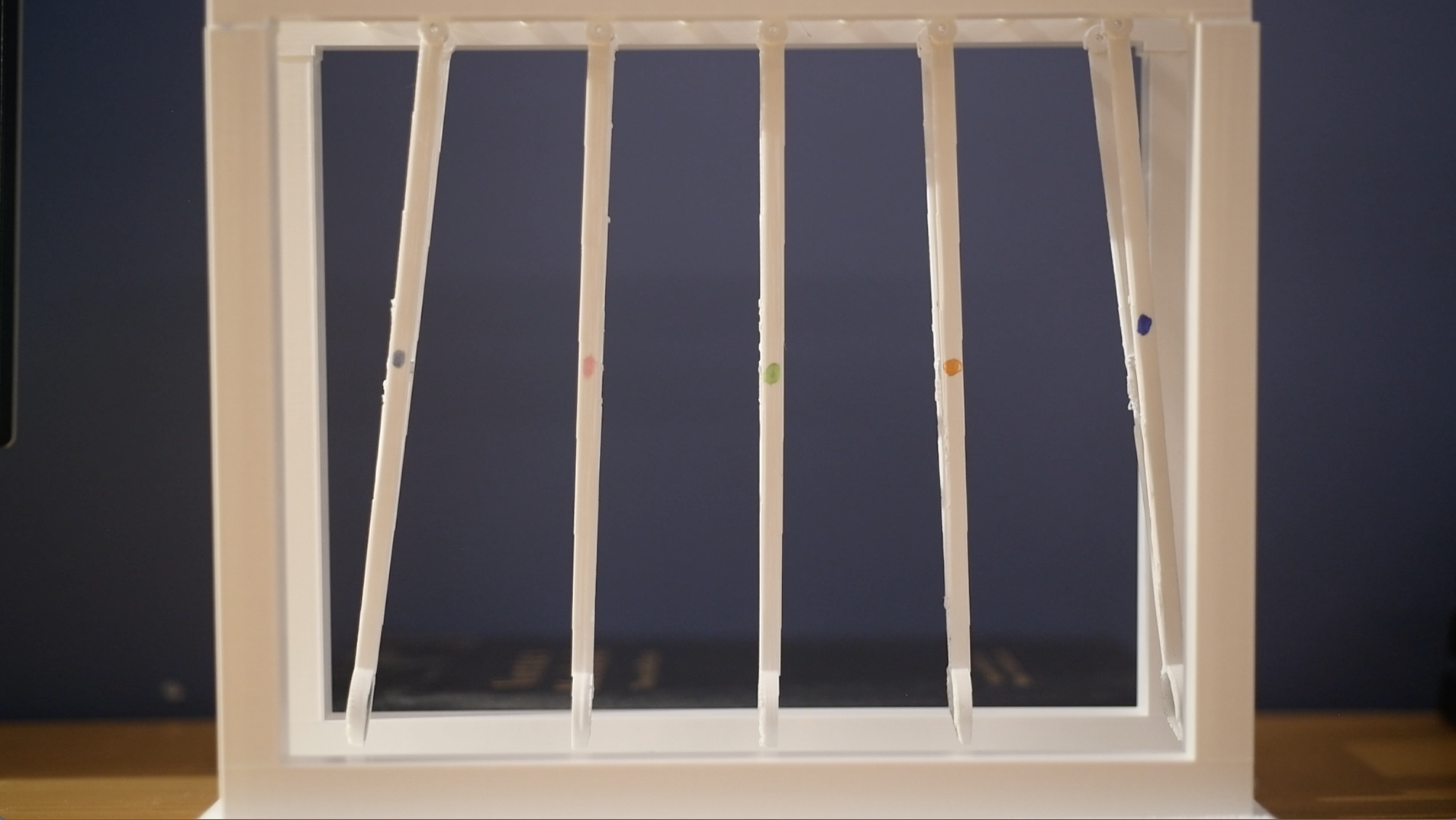}\hfill
\caption{The 3D printed \ac{MNC}. Here we show the 5 magnet model. Color dots on the \ac{MNC} arms are used for the motion tracking.}
\label{fig:MNCsketch}
\end{figure}

In this section we describe the \ac{MNC} apparatus and the experimental protocol used in this study.
Experiments were carried out on a self-made, 3D-printed frame (Fig.~\ref{fig:MNCsketch}) that holds $N=2$–$5$ identical pendulums, each a thin rod terminating in a circular enclosure containing 
axially magnetized N45 neodymium disc magnet (\SIlist{15.9;1.6}{mm}). Magnets are mounted in
\textbf{alternating polarity}, $s_k=(-1)^k$, so that neighboring magnets repel and next nearest
neighbors attract. Pivot spacing is adjustable over \SIrange{2}{10}{cm}. The pendulum motion is filmed at high
frame rate using a Nikon D850 camera and digitized with Tracker software~\cite{Tracker2026} to compute the angle $\theta_k(t)$ of every pendulum. In each test, the pendulums 
are released from rest at a measured held angle, $\theta_k(0)=\theta_k^{\mathrm{held}}$,
$\dot\theta_k(0)=0$.

To develop the models every relevant parameter is measured independently
(Table~\ref{tab:constants}). Both damping coefficients are obtained from single-pendulum free
decays (\S\ref{sec:damping-results}); the magnet strength $\Imag$, once determined, is transferred
unchanged between configurations.

\begin{table}[H]\centering
\begin{tabular}{lp{0.42\linewidth}l}
\toprule
quantity & method & value\\
\midrule
magnet radius $R$ & calipers (\SI{15.9}{mm} diameter) & \SI{7.95}{mm}\\
magnet thickness $L$ & calipers  & \SI{1.6}{mm}\\
length / center of mass $l,\lcom$ & balance against a known mass & \SI{0.173}{m} / \SI{0.104}{m}\\
mass $M$ & scale & \SI{7.0}{g}\\
moment of inertia $I$ & bare-swing period, $I=Mg\lcom T^2/4\pi^2$ & \SI{9.25e-5}{kg.m^2}\\
natural frequency $f_0$ & each pendulum's free swing from camera data& ${\approx}\SI{1.45}{Hz}$\\
magnet strength $\Imag$ & Hall-probe axial field map, all 5 magnets (\S\ref{sec:hallprobe}) & $\SI{1160\pm34}{A}$\\
\bottomrule
\end{tabular}
\caption{Independently measured constants. The value of $\Imag$ is the direct Hall-probe
measurement of \S\ref{sec:hallprobe} (mean $\pm$ standard error over five magnets); the
closed-loop predictions of \S\ref{sec:validation} use the value fitted once to a three-pendulum
recording (${\approx}\SI{1120}{A}$), which agrees with it.}
\label{tab:constants}
\end{table}

\subsection{Theoretical model}
\label{sec:model}
The angular oscillations of each pendulum are determined by the torques due to the magnetic interactions between pendulums, gravity, as well as air drag for each pendulum. 

We investigate two models for the magnet's behavior: (i) magnet as a current loop and a (ii)  magnet as a dipole
\subsubsection{A magnet as a current loop}
A uniform magnetization $\mathbf M$ is equivalent to bound currents $\mathbf J_b=\nabla\times
\mathbf M$ (volume) and $\mathbf K_b=\mathbf M\times\hat{\mathbf n}$ (surface). For a uniformly
 axially-magnetized disk ($\mathbf M=M\hat{\mathbf z}$, constant), $\nabla\times\mathbf M=0$, so
there is no volume current; on the rim, $\mathbf K_b=M\,\hat{\mathbf z}\times\hat{\boldsymbol\rho}
=M\,\hat{\boldsymbol\varphi}$, there is a current circulating around the edge. For thickness $L\ll R$ this
surface sheet reduces to a single loop of current
\begin{equation}
\;\Imag=M\,L\;.
\end{equation}
The exact equivalent of the magnet is a surface current \emph{sheet} of height $L$; the single
thin loop is its $L\ll R$ limit. For the disks in this study  (see Table I) we have $L=\SI{1.6}{mm}$ and $R=\SI{8}{mm}$, so this is a reasonable 
approximation. This approximation is also checked below against the closed-form finite-cylinder force~\cite{vokoun2009,ravaud2010}.

The Biot–Savart law for a loop $C$ of radius $R$ in the $z=0$ plane,
$\mathbf B(\mathbf r)=\tfrac{\mu_0\Imag}{4\pi}\oint_C d\boldsymbol\ell'\times(\mathbf r-\mathbf
r')/|\mathbf r-\mathbf r'|^3$, is evaluated at a field point $(\rho,0,z)$ by parametrizing the
loop with $\varphi'$ and substituting $\varphi'=\pi+2\psi$; every term collapses onto the complete
elliptic integrals\cite{Jackson1999}
\begin{equation}
K(m)=\int_0^{\pi/2}\!\frac{d\psi}{\sqrt{1-m\sin^2\psi}},\qquad
E(m)=\int_0^{\pi/2}\!\sqrt{1-m\sin^2\psi}\;d\psi,\quad
m=\frac{4R\rho}{R^{2}+\rho^{2}+z^{2}+2R\rho}.
\end{equation}
With $\alpha=R^{2}+\rho^{2}+z^{2}-2R\rho$, $\beta=R^{2}+\rho^{2}+z^{2}+2R\rho$,
\begin{equation}
\;B_z=\frac{\mu_0\Imag}{2\pi\sqrt\beta}\!\left[\frac{(R^{2}-\rho^{2}-z^{2})E(m)}{\alpha}+K(m)\right],\quad
B_\rho=\frac{\mu_0\Imag\,z}{2\pi\rho\sqrt\beta}\!\left[\frac{(R^{2}+\rho^{2}+z^{2})E(m)}{\alpha}-K(m)\right]\;
\label{eq:loopfield}
\end{equation}
with $B_\rho\to0$ on the axis. Every component is linear in $\Imag$ and exact at all distances —
in particular, at the sub-diameter separations the cradle operates at.

\subsubsection{The far-field (dipole) limit}
For large distances between magnets ($r=\sqrt{\rho^2+z^2}\gg R$),  we find that $m\to0$, $K,E\to\pi/2$, and~\eqref{eq:loopfield} reduce to the
field of a point dipole of moment $\mathbf m=\Imag\,\pi R^{2}\,\hat{\mathbf z}$, with
\begin{equation}
B_z^{\text{axis}}\to \tfrac{\mu_0}{4\pi}\tfrac{2m}{r^{3}}
\end{equation}, and the force between two dipoles
reduces to the familiar $1/r^4$ law used in most prior magnetic-pendulum work~\cite{russell1997,cuevas2009}. With $\Imag\approx\SI{1120}{A}$, $m\approx\SI{0.22}{A.m^2}$. Figure~\ref{fig:loopdipole} quantifies where the two models diverge: the dipole over-predicts the force
by ${\sim}86\%$ at \SI{2}{cm} separation, falling to ${\sim}5\%$ only by \SI{8}{cm}. When the
cradle is operated over this range, the exact loop field of~\eqref{eq:loopfield}  is required for an accurate predictive model.

\fg[0.72\linewidth]{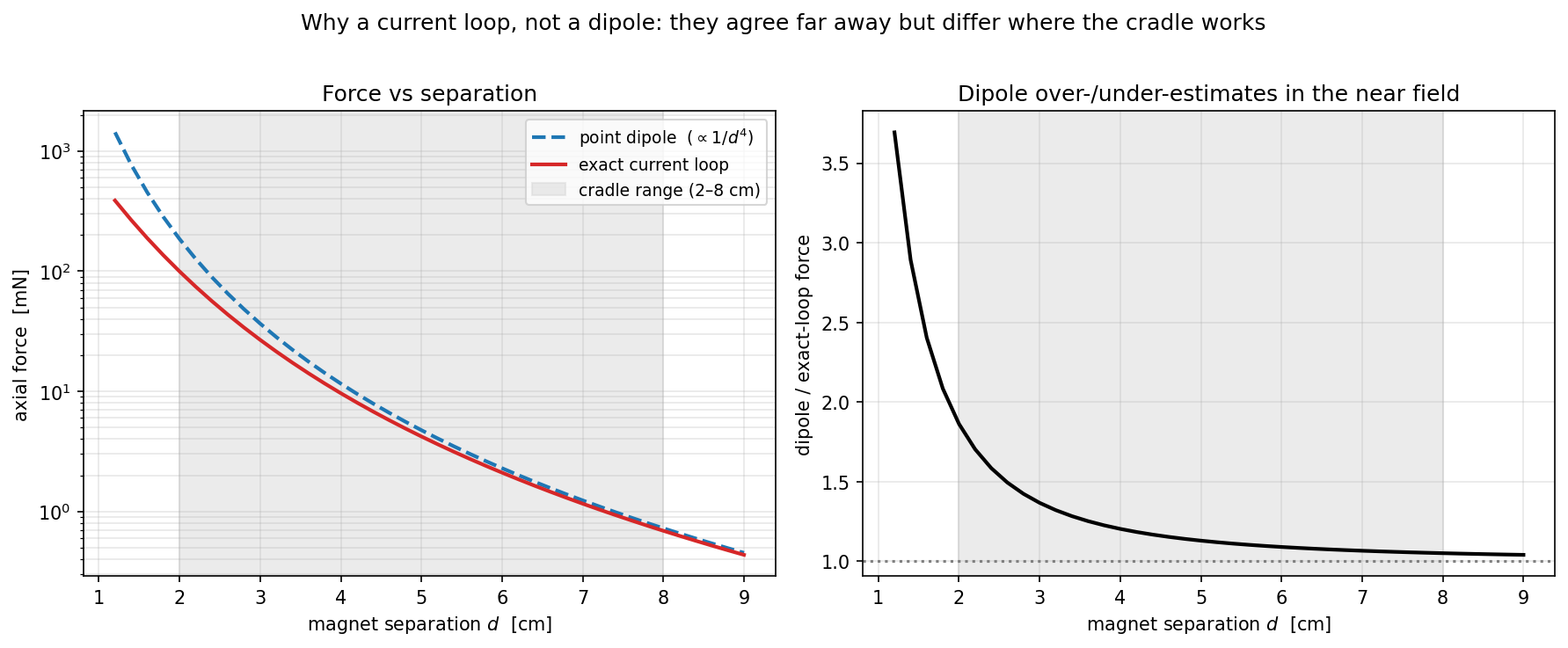}{Force between two magnets:  current-loop
model vs.\ the point-dipole approximation. The dipole over-predicts by ${\sim}86\%$ at
\SI{2}{cm}, converging to the loop result only beyond ${\sim}\SI{8}{cm}$ (shaded band = the
cradle's operating range).\label{fig:loopdipole}}

\subsubsection{Inter-magnet force, torque, and the \texorpdfstring{$\Imag^2$}{I\_mag\textasciicircum 2} scaling}
A current carrying loop exposed to  an external field experiences the Lorentz (motor) force; loop
$i$ in loop $j$'s field feels
\begin{equation}
\;\mathbf F_{i\leftarrow j}=\oint_{\text{loop }i}\Imag\,d\boldsymbol\ell\times\mathbf
B_j(\mathbf r')\;,
\end{equation}
evaluated numerically by the trapezoidal rule around the loop. Since $\mathbf B_j\propto\Imag$ and
the integrand carries a further factor of $\Imag$, the force — and hence the torque — scales
exactly as
\begin{equation}
\mathbf F,\ \tau \;\propto\; \Imag^{2}.
\end{equation}
With alternating orientation $s_k=(-1)^k$, the force carries a factor $s_is_j$: $-1$ (repel) for
neighbors, $+1$ (attract) for next-nearest neighbors. Pendulum $k$ hangs from pivot $x=k\,d_p$
with magnet position $\mathbf r^{\mathrm{rel}}_k=(-l\sin\theta_k,\,l\cos\theta_k)$, giving the
planar torque about the pivot, $\tau^{\mathrm{mag}}_{k\leftarrow j}\propto\Imag^2$. Because this
torque is $\Imag^2$ times a function of angles alone, we tabulate the unit-current torque
$\tau^{\mathrm{unit}}_{kj}(\theta_k,\theta_j)$ once per spacing and rescale by $\Imag^2$ at run
time.

\subsubsection{Equations of motion and damping}
\label{sec:damping}

Based on the above observations  the governing equations for the motion of the pendulums are expressed as: 
\begin{equation}
\boxed{
I_k\,\ddot\theta_k=\underbrace{\Imag^{2}\!\sum_{j\ne k}\tau^{\mathrm{unit}}_{kj}(\theta_k,\theta_j)}_{\text{magnetic interactions}}
-\underbrace{Mg\,\lcom\sin\theta_k}_{\text{gravity}}
-\underbrace{b\,\dot\theta_k}_{\text{linear drag}}
-\underbrace{c\,\dot\theta_k|\dot\theta_k|}_{\text{quadratic drag}}, \;\;k=1,\cdots N_k
}
\label{eq:eom}
\end{equation}
where $N_k$ denotes the total number of pendulums.

\subsection{Parameter identification by Optimizing a Discrete Loss}
\label{sec:odil}

The model parameters of the cradle can be identified by using  a video of the experimental apparatus. One approach to find the parameters is by 
optimization: initialize the model parameters integrate the equations of motion
forward in time, compare the simulated angles with the measured ones, adjust the parameters, repeat\cite{ribeiro2020smoothness}.
This works well when the motion is gentle. It works badly when the motion is chaotic, because
then a tiny error in the parameter makes the simulated trajectory wander off from the measured
one after a few swings, and the optimization process diverges.

In \ac{ODIL}~\cite{karnakov2024odil,karnakov2023flow} the angles of all pendulums at \emph{all}
time steps are treated as unknowns, \emph{in addition} to the
physical constants. These unknowns are identified via an optimization requiring that the equations of motion hold as well as possible at every time step and
the simulated angles pass as close as possible to the measured ones. These requirements are
written as sums of squares, the \emph{discrete loss}
$\mathcal L$ which is then minimized. Schematically,
\begin{equation}
\boxed{\;\mathcal L=\underbrace{\sum_{\text{time steps}}\big(\text{equation-of-motion error}\big)^{2}}_{\text{physics}}
\;+\;\underbrace{\sum_{\text{measurements}}\big(\text{simulated}-\text{measured}\big)^{2}}_{\text{data}}\;}
\label{eq:odil-general}
\end{equation}
 If there are no data and the constants are known, the minimum of
$\mathcal L$ is zero and the minimizer is exactly the solution a time-stepping scheme would
produce. In this case  \ac{ODIL} must reproduce a conventional simulation to round-off, a useful self-check.
If there are data the physics term only nudges the trajectory
towards obeying the equations of motion.  The data term at the minimum is a
direct measure of how well the \emph{model} can explain the data \emph{throughout the recording}. This allows for the two models using the dipole and current-loop magnets to be compared on equal terms.

\subsubsection{The cradle model used for identification}
\label{sec:odil-model}
We use a simplified version of~\eqref{eq:eom}
with $N=3$ pendulums: each is a point mass at the end of a massless string of length $l$, the
magnet sits at the mass, damping is linear, and the magnets keep their orientation as the
pendulums swing. Measuring lengths in units of $l$ and
time in units of $\sqrt{l/g}$, the angle
$\theta_k$ of pendulum $k$ obeys
\begin{equation}
\boxed{\;\ddot\theta_k=-\sin\theta_k\;-\;\gamma\,\dot\theta_k\;+\;\kappa\,q_k(\theta_1,\theta_2,\theta_3)\;},\qquad k=1,2,3,
\label{eq:reduced}
\end{equation}
where the three terms are gravity, damping and the magnetic torque. The two dimensionless
constants,
\begin{equation}
\gamma=\frac{b}{Ml^{2}\sqrt{g/l}}\quad\text{(damping)},\qquad
\kappa=\frac{3\mu_0\,m^{2}}{4\pi\,Mg\,l^{4}}\quad\text{(magnet strength)},\qquad m=\Imag\,\pi R^{2},
\label{eq:kappa}
\end{equation}
will be identified by \ac{ODIL}. Magnets are mounted with alternating
polarity $s_k=(-1)^{k}$, so neighbors repel and the two end magnets attract weakly. Pendulum
$k$ hangs from $x=k\,d_p/l$, so its magnet sits at
$\mathbf r_k=(k\,d_p/l+\sin\theta_k,\,-\cos\theta_k)$.

The magnetic torque $q_k$ is where the two magnet models differ. Both are written as an
interaction energy $\tilde U$ between pairs of magnets; the torque on pendulum $k$ is then
simply minus the slope of that energy with respect to $\theta_k$,
\begin{equation}
q_k=-\frac{\partial\tilde U}{\partial\theta_k},\qquad \tilde U=\sum_{\text{pairs }i<j}\tilde U_{ij},
\label{eq:qk}
\end{equation}
which we evaluate by \ac{AD} in JAX so that both models run through identical
code.

\paragraph{Model D (point dipole).}
Treating each magnet as a point dipole, the pair energy depends only on the distance $r_{ij}$
between the two magnet centers and on how far the line joining them, $\Delta x_{ij}$, is from the
magnets' common axis:
\begin{equation}
\tilde U^{\rm D}_{ij}=\frac{s_is_j}{3\,r_{ij}^{3}}\Big(1-\frac{3\,\Delta x_{ij}^{2}}{r_{ij}^{2}}\Big).
\label{eq:Udip}
\end{equation}
For two magnets side by side on the axis ($\Delta x=r$) this gives a force of magnitude
$2/r^{4}$ in these units, i.e.\ the familiar far-field repulsion $6\mu_0m^{2}/(4\pi r^{4})$
between antiparallel dipoles. The energy grows without bound as $r\to0$, so a tiny softening
distance ($r^{2}\to r^{2}+\epsilon^{2}$, $\epsilon=10^{-3}$) is added to keep intermediate
optimizer iterates finite.

\paragraph{Model L (exact current loop).}
Here each magnet is a circular loop of wire of radius $a=R/l$ carrying a current
($m/\pi a^{2}$) that gives it the same magnetic moment as the dipole. The energy of two current loops
is, $U=-I_1I_2\mathcal M_{12}$, where the mutual inductance $\mathcal M_{12}$
is Neumann's double line integral around both loops. In SI units,
\begin{equation}
\tilde U^{\rm L}_{ij}=-\frac{s_is_j}{3\pi^{2}a^{2}}\oint\!\!\oint
\frac{\cos(\varphi_1-\varphi_2)}{\big|\boldsymbol\rho_i(\varphi_1)-\boldsymbol\rho_j(\varphi_2)\big|}\,d\varphi_1\,d\varphi_2,
\label{eq:Uloop}
\end{equation}
where $\boldsymbol\rho_i(\varphi)$ is the point on loop $i$ at angle $\varphi$ around its rim and
$\cos(\varphi_1-\varphi_2)$ is the dot product of the two tangent directions. Once the pendulums
swing the two loops are parallel but no longer share an axis, and there is no closed form; we
simply place $Q=24$ equally spaced points around each loop and replace the double integral by a
double sum over the $24\times24$ pairs of points. Because the integrand is smooth and periodic
this ``trapezoidal rule'' is extremely accurate: with $Q=24$ we reproduce the exact
elliptic-integral result for two coaxial loops~\cite{vokoun2009,ravaud2010} to twelve digits.
When the loop radius $a$ is small compared with the distance $r$ between magnets, Model L reduces
to Model D with a correction of order $(a/r)^{2}$—that is why both models can share the same
constant $\kappa$. When $a$ is comparable to $r$, as it is at the closest approach in the cradle,
the two models differ substantially (Fig.~\ref{fig:loopdipole} and \S\ref{sec:odil-results}).

The numerical values used in \S\ref{sec:odil-results}-$\kappa=5\times10^{-5}$, $\gamma=0.02$,
pivot spacing $d_p/l=0.2$ and $a=R/l\in\{0.0125,0.025,0.05\}$-are of the same order as in
the apparatus of Table~\ref{tab:constants} ($R/l=0.046$, $d_p/l=0.23$ at \SI{4}{cm} spacing).

\subsubsection{Optimization using a discrete loss}
\label{sec:odil-discrete}

The principle of \ac{ODIL} requires that we divide the time interval $[0,T]$ into $N_t$ equal steps of length $h$ and call the (unknown)
angles at step $n$ $\boldsymbol\theta^{n}=(\theta^{n}_1,\theta^{n}_2,\theta^{n}_3)$. Replacing the
second derivative in~\eqref{eq:reduced} by a  central difference and the first
derivative by a centered difference gives, at every interior time step, the \emph{residual}
\begin{equation}
\boxed{\;\mathbf R^{n}=\underbrace{\frac{\boldsymbol\theta^{n+1}-2\boldsymbol\theta^{n}+\boldsymbol\theta^{n-1}}{h^{2}}}_{\ddot{\boldsymbol\theta}}
+\sin\boldsymbol\theta^{n}
+\gamma\,\underbrace{\frac{\boldsymbol\theta^{n+1}-\boldsymbol\theta^{n-1}}{2h}}_{\dot{\boldsymbol\theta}}
-\kappa\,\mathbf q(\boldsymbol\theta^{n})\;},\qquad n=1,\dots,N_t-1,
\label{eq:stencil}
\end{equation}
Two more residuals impose the release from rest at the held angles $\boldsymbol\theta_0$:
$\mathbf R_{\rm ic,0}=\boldsymbol\theta^{0}-\boldsymbol\theta_0$ fixes the starting angles and
$\mathbf R_{\rm ic,1}=(\boldsymbol\theta^{1}-\boldsymbol\theta^{0})/h-\tfrac h2\,\ddot{\boldsymbol\theta}(0)$
fixes the starting velocity to zero (the second term is the small displacement gravity and
magnets produce during the first half step). Finally, each measurement---pendulum $k_m$ observed
at time step $n_m$ with value $\theta^{\rm obs}_m$---contributes the residual
$\theta^{n_m}_{k_m}-\theta^{\rm obs}_m$. The discrete loss is the weighted sum of all these
residuals squared:
\begin{equation}
\boxed{\;\mathcal L=h\sum_{n=1}^{N_t-1}\|\mathbf R^{n}\|^{2}
+\|\mathbf R_{\rm ic,0}\|^{2}+\|\mathbf R_{\rm ic,1}\|^{2}
+\sum_{m=1}^{N_d}\big(\theta^{n_m}_{k_m}-\theta^{\rm obs}_{m}\big)^{2}\;}
\label{eq:odil-loss}
\end{equation}
The unknowns are all the $\boldsymbol\theta^{n}$
and, in the inverse problem, the constants to be identified. We optimize $\log\kappa$,
$\log\gamma$ and $\log a$ rather than the constants themselves; this keeps them positive
automatically, and one optimizer step changes them by a percentage rather than by an amount.

For the \emph{forward} problem (constants known, no data) $\mathcal L=0$ is attainable and the
minimizer is exactly the trajectory one gets by solving~\eqref{eq:stencil} for
$\boldsymbol\theta^{n+1}$ and marching forward in time,
\begin{equation}
\boldsymbol\theta^{n+1}=\Big(\frac1{h^{2}}+\frac{\gamma}{2h}\Big)^{-1}\Big[\frac{2\boldsymbol\theta^{n}-\boldsymbol\theta^{n-1}}{h^{2}}
+\frac{\gamma}{2h}\boldsymbol\theta^{n-1}-\sin\boldsymbol\theta^{n}+\kappa\mathbf q(\boldsymbol\theta^{n})\Big].
\label{eq:verlet}
\end{equation}
We use this recursion in two ways: as the check that the optimizer has really found the minimum,
and as an inexpensive  way to produce a starting trajectory. Synthetic measurements for the tests in
\S\ref{sec:odil-results} are generated with a high-order integrator (DOP853, relative tolerance
$10^{-11}$) of the continuous equations, so that the $h^{2}$ discretization error of~\eqref{eq:stencil} is present in every inverse test just as it would be with real data.

\subsubsection{Minimization of the loss}
\label{sec:odil-newton}
$\mathcal L$ is a sum of squares of residuals $\mathbf r(\mathbf z)$, where $\mathbf z$ collects
all unknowns. We use the Gauss--Newton method which linearizes the residuals
about the current guess, $\mathbf r(\mathbf z+\delta)\approx\mathbf r+\mathbf J\delta$ with
$\mathbf J=\partial\mathbf r/\partial\mathbf z$ the Jacobian matrix, and choose the step $\delta$
that minimizes the linearized sum of squares,
\begin{equation}
\big(\mathbf J^{\!\top}\mathbf J+\lambda\,\mathbf D\big)\,\delta=-\mathbf J^{\!\top}\mathbf r,
\qquad \mathbf z\leftarrow\mathbf z+\delta .
\label{eq:gn}
\end{equation}
The extra term $\lambda\mathbf D$ (Levenberg--Marquardt damping, $\mathbf D$ the diagonal of
$\mathbf J^{\!\top}\mathbf J$) shortens the step when the linearization is poor: $\lambda$ is
decreased after every successful step and increased after every step that fails to lower
$\mathcal L$. Near the solution the iteration converges quadratically.

The Jacobian is large ($\sim6000\times6000$ for $N_t=2000$) but sparse: residual
$\mathbf R^{n}$ depends only on $\boldsymbol\theta^{n-1},\boldsymbol\theta^{n},\boldsymbol\theta^{n+1}$,
so each row of $\mathbf J$ has just nine non-zeros (plus one per identified constant). We build it
exactly with nine forward-mode automatic-differentiation passes, using the trick that time steps
$n$, $n+3$, $n+6,\dots$ never share a residual and can therefore be differentiated together
(Appendix~\ref{app:odil}), store it as a sparse matrix, and solve~\eqref{eq:gn} with a sparse LU
factorization. One iteration then costs a time proportional to $N_t$: about \SI{0.05}{s} for
Model D and \SI{0.6}{s} for Model L (whose torque involves the $24\times24$ double sum for each
pair of magnets at each time step) on a laptop. This ``Newton mode'' of \ac{ODIL} is what we use for all
results. For comparison, we also minimize the same loss with a gradient-based method of the kind
used to train neural networks (Adam, followed by L-BFGS); \S\ref{sec:odil-results} shows why the
Newton mode is preferred here.

\paragraph{Starting guesses.}
For the forward problem
we start from the constant trajectory $\boldsymbol\theta^{n}=\boldsymbol\theta_0$ on a coarse grid
with $N_t/4$ steps, converge, interpolate the solution onto a grid twice as fine, converge again,
and repeat, a one-way version of the multigrid idea of~\cite{karnakov2023flow}. When solving a
family of forward problems (several magnet radii) the previous solution is the starting point of
the next. For the inverse problem we start from the trajectory obtained by running the recursion~\eqref{eq:verlet} with the \emph{guessed} constants; the measurements themselves are deliberately
not used to build the starting trajectory, for a reason shown in \S\ref{sec:odil-results}.
Algorithm~\ref{alg:odil} summarizes the procedure.

\begin{algorithm}[H]
\caption{\ac{ODIL} identification for the magnetic cradle.}
\label{alg:odil}
\begin{algorithmic}[1]
\Require magnet model (D or L); known constants; held angles $\boldsymbol\theta_0$; time grid $(T,N_t)$; measurements (none for a forward problem); initial guess for the constants to be identified
\State trajectory $\gets$ recursion~\eqref{eq:verlet} with the guessed constants
\State $\lambda\gets10^{-3}$
\Repeat
  \State evaluate all residuals~\eqref{eq:stencil}--\eqref{eq:odil-loss}, with the torque from~\eqref{eq:qk} and the energy of Model D~\eqref{eq:Udip} or Model L~\eqref{eq:Uloop}
  \State assemble the sparse Jacobian $\mathbf J$ (nine \ac{AD} passes for the trajectory, one per identified constant)
  \State solve~\eqref{eq:gn} for the step $\delta$
  \State \textbf{if} the step lowers $\mathcal L$: accept it and decrease $\lambda$; \textbf{else} reject it and multiply $\lambda$ by 4
\Until{$\mathcal L$ is below tolerance or no step is accepted}
\Ensure trajectory, identified constants, \ac{RMS} data misfit, physics residual
\end{algorithmic}
\end{algorithm}

\label{sec:odil-protocol}
Three uses of Algorithm~\ref{alg:odil} are reported in \S\ref{sec:odil-results}.
\emph{Forward consistency:} no constants identified, no data; the \ac{ODIL} trajectory must agree
with the recursion~\eqref{eq:verlet} to round-off, and differs from the DOP853 solution only by
the $h^{2}$ error of the scheme. \emph{Identification:} data generated with Model D and known
$(\kappa,\gamma)$ are fitted by Model D starting from a deliberately wrong guess, for several
observation patterns and noise levels, to see how accurately the constants come back.
\emph{Model discrimination:} data generated with the finite-size Model L are fitted by
(i) Model D for $(\kappa,\gamma)$, (ii) Model L with the radius known, and (iii) Model L for
$(\kappa,\gamma,a)$. If the point dipole were an adequate description, fit (i) would return the
true $\kappa$ and a misfit at the noise level. How far it misses in $\kappa$, equivalently in
$\Imag\propto\sqrt\kappa$, and in the leftover misfit to noise free data, measures the
model-form error of the far-field approximation.

\paragraph{Calibration on the measured trajectories.}
For the experimental data, the same algorithm is run with the rigid-rod model of~\eqref{eq:eom}
rather than the point-mass twin: the magnets sit a distance $l$ from the pivot and turn with the
rod, and the time unit is $1/\omega_0$ with $\omega_0=2\pi f_0$, so that
$\kappa=3\mu_0m^{2}/(4\pi I\omega_0^{2}l^{3})$ and $\gamma=b/(I\omega_0)$. Four practical
differences matter. (i)~The recordings start while the first pendulum is still held, so the frames
before release are dropped and the initial angles and velocities of all pendulums are left free, no
release-from-rest assumption is made. (ii)~Each rod may have a slightly different natural
frequency and each tracked angle a small calibration offset, so per-pendulum frequency factors
$f_k$ (prior $1\pm3\%$) and offsets $\delta\theta_k$ (prior $0\pm1^\circ$) are added as
nuisance unknowns. The damping enters through $\gamma=b/(I\omega_0)$ and $\gamma_2=c/I$ (the
quadratic term of~\eqref{eq:eom}), either of which can be held at its directly measured value or
released as an unknown; a third, optional term $-\gamma_f\,\overline{\dot\theta}$ acting on the mean
angular velocity of the row is used in \S\ref{sec:odil-damping} to test for losses through the common
motion of all pendulums. (iii)~The physics weight in~\eqref{eq:odil-general} is increased in stages
(continuation from $w=1$ to $w=10^{4}$): the first, ``soft'' stage lets the trajectory follow the
data while the constants adjust; the last, ``hard'' stage forces the trajectory to obey the model,
so that its remaining misfit is the genuine predictive error of the calibrated model. (iv)~Every
video frame is a measurement of all $N$ angles, so the observation set is dense, and the solver
grid uses two steps per frame.

%

\section{Results}
\label{sec:results}
Here we present an analysis of the linear modes to show the agreement of our method with existing works. We then describe results from the inverse modeling process and compare the two models for the magnets.
\subsection{Linear normal modes}
\label{sec:linmodes}

For a classical \ac{NC}, small-amplitude approximations $\theta\approx A\cos\omega_0 t$ with
$\omega_0=\sqrt{Mg\lcom/I}$, energy $E=\tfrac12 Mg\lcom A^{2}$, and the drag torque
$\tau_{\rm drag}=b\dot\theta+c\dot\theta|\dot\theta|$ removes energy at a rate
$\Delta E=-b\pi A^2\omega_0-\tfrac83 cA^3\omega_0^2$ per cycle, giving the amplitude-decay equation
\begin{equation}
\boxed{\;\frac{dA}{dt}=-\alpha A-\beta A^{2},\qquad \alpha=\frac{b}{2I},\quad
\beta=\frac{4c\,\omega_0}{3\pi I}\;}
\label{eq:decay}
\end{equation}
When we model the drag using only the linear term we obtain a constant fractional decay rate. The  quadratic drag adds a correction that grows with amplitude and dominates above
$|\dot\theta|=b/c$. The equation of motion~\eqref{eq:eom} is integrated with a differentiable
solver (\texttt{diffrax}/JAX). In turn the model  parameters are fitted with \texttt{scipy}'s trust-region
least-squares.

Writing $\boldsymbol\theta=\boldsymbol\theta^{*}+\delta\boldsymbol\theta$ about the static
equilibrium $\boldsymbol\tau(\boldsymbol\theta^{*})=\mathbf 0$ and Taylor-expanding the total
torque to first order gives $I\,\delta\ddot{\boldsymbol\theta}=-\mathbf K\,\delta\boldsymbol
\theta$, with stiffness matrix
\begin{equation}
K_{kj}=Mg\lcom\cos\theta^{*}_k\,\delta_{kj}\;-\;\Imag^{2}\,\frac{\partial\,\tau^{\mathrm{unit}}_{k}}
{\partial\theta_j}\Big|_{\boldsymbol\theta^{*}} .
\end{equation}
Seeking solutions $\delta\boldsymbol\theta=\mathbf v_n e^{i\omega_n t}$ gives the generalized
eigenvalue problem
\begin{equation}
\boxed{\;\mathbf K\,\mathbf v_n=I\,\omega_n^{2}\,\mathbf v_n,\qquad f_n=\frac{\omega_n}{2\pi}\;}
\label{eq:eig}
\end{equation}
An $N\times N$ symmetric $\mathbf K$ has $N$ real eigenvalues, hence $N$ normal modes for $N$
pendulums. The lowest mode is the in-phase mode, $\mathbf v_1\approx(1,\dots,1)$: for vertical rods the
magnet spacings do not change and $\omega_1=\omega_0$; the static offsets $\theta^*_k$ shift it by
$O(\theta^{*2})$, less than $1\%$ for the present rig. In this mode the magnetic interaction is
not loaded, just as touching balls swinging together are not compressed in a classical cradle.
Higher modes compress the magnets and are magnetically stiffened, raising their frequency.
\label{sec:symmetry}%
The center pendulum sits at a node of every antisymmetric mode and shows only the
$\lceil N/2\rceil$ symmetric modes — 3 of 5 for $N=5$, 2 of 3 for $N=3$ — a prediction confirmed
directly in the per-pendulum Fourier spectra (\S\ref{sec:modes-results}).

The eigenfrequencies of~\eqref{eq:eig} are exact in the $A\to0$ limit. At  the small amplitudes
of the recordings analyzed here ($\lesssim7^\circ$) the finite-amplitude corrections to the
gravitational and magnetic restoring torques shift them by less than $1\%$, and they are not
treated further in this paper.

\subsection{Direct measurement of the magnet strength}
\label{sec:hallprobe}
Throughout \S\ref{sec:model}--\S\ref{sec:linmodes} the magnet strength $\Imag$ was treated as a
constant to be measured. We have  made that measurement
directly. We used  a Hall probe that was swept along the axis of each of the five magnets used in the cradle,
recording the axial field $B_z$ from \SI{0.2}{mm} to \SI{57}{mm} from the magnet face
(Fig.~\ref{fig:hallprobe}). Each sweep is fit to the \emph{exact} finite current-sheet field of a
uniformly magnetized cylinder of height $L$ and radius $R$ (the surface-current generalization of~\eqref{eq:loopfield}, with surface current density $K=\Imag/L$),
\begin{equation}
B_z(z)=\frac{\mu_0 K}{2}\left[\frac{z+L/2}{\sqrt{R^2+(z+L/2)^2}}-\frac{z-L/2}{\sqrt{R^2+(z-L/2)^2}}\right],
\end{equation}
rather than the thin-loop limit, since several probe positions are closer to the magnet than its
own thickness, where the thin-loop approximation is least reliable. The five independent fits give
\begin{equation}
\Imag = 1131,\ 1228,\ 1146,\ 1055,\ 1242\ \text{A}\quad\Rightarrow\quad
\boxed{\;\Imag = \SI{1160\pm34}{A}\;}
\end{equation}
(mean $\pm$ standard error; magnet-to-magnet spread 6.6\%), which agrees with the value fitted to the multi-pendulum dynamics
(${\approx}\SI{1120}{A}$; the \ac{ODIL} calibrations of \S\ref{sec:odil-exp} give 1112--\SI{1151}{A}) to within $4\%$, about one standard error. Hence, the two independent
determinations of $\Imag$, one magnetostatic and one dynamical, agree. The implied remanent field
$\mu_0\Imag/L\approx\SI{0.91}{T}$ is lower than the N45 datasheet remanence
(\SIrange{1.32}{1.35}{T}). Because the pendulum dynamics independently give the same moment, the difference is unlikely to
be an artifact of the field fit; the most likely explanation is that the magnets are of a lower
grade than specified.

\fg[0.66\linewidth]{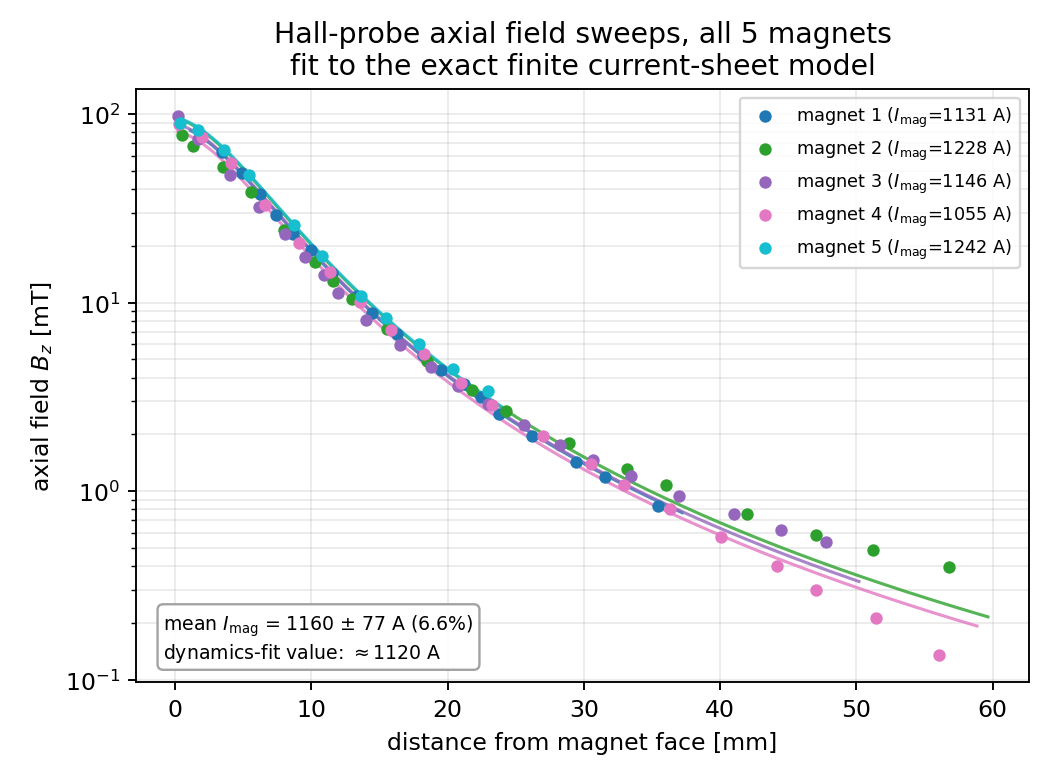}{Axial field sweeps for all five magnets used in
the cradle, each fit independently to the exact finite current-sheet model. The five fitted values
of $\Imag$ (legend) agree with each other to $6.6\%$ and with the dynamics-fit value used
throughout the predictions of \S\ref{sec:validation} to within $4\%$.\label{fig:hallprobe}}

\subsection{Damping: from an indirect estimate to a direct measurement}
\label{sec:damping-results}
A small amplitude bare pendulum free decay is a clean exponential on a log-amplitude plot,
consistent with linear drag. Fitting its slope gives $b\approx\SI{1.24e-5}{N.m.s}$. This value,
however, underdamps the multi-pendulum cradle and the
gap between model and data grows with amplitude. Hence, some additional, amplitude-dependent
dissipation is implied  \emph{by the data}. We subsequently recorded a dedicated
large-amplitude (${\sim}37^\circ$) single-pendulum free decay and jointly fit both terms of~\eqref{eq:decay} to that one record, resolving $b$ and $c$ self-consistently:
\begin{equation}
b = \SI{2.65e-5}{N.m.s},\qquad c = \SI{2.72e-6}{N.m.s^2},\qquad \text{envelope \ac{RMS} } \SI{0.71}{\degree}.
\end{equation}
This directly measures the quadratic term from an isolated pendulum, rather than inferring it
indirectly from the multi pendulum residual. The linear coefficient of this joint fit is, however,
twice the small-amplitude value; a single $(b,c)$ pair does not describe both decays, so the
damping law of~\eqref{eq:eom} is itself only approximate. The revised
$b$ gives a damping ratio $\zeta=b/(2I\omega_0)\approx0.016$, with a
$1/e$ amplitude decay time of $\sim\SI{7}{s}$. Propagating the refined pair through the closed-loop
multi-pendulum predictions of \S\ref{sec:validation} changes the reported \ac{RMS} errors by only a few
hundredths of a degree, confirming that the magnetic coupling, not damping, governs the cradle's
dynamics. We checked candidate
physical mechanisms for the quadratic term: ordinary aerodynamic drag is ruled out and eddy-current
damping at these field strengths is very weak. We therefore treat $c$ as an \emph{effective} interaction loss term, most plausibly
representing  out-of-plane or frame motion at the close magnetic approaches, rather than
a proven aerodynamic effect.

\fg[0.68\linewidth]{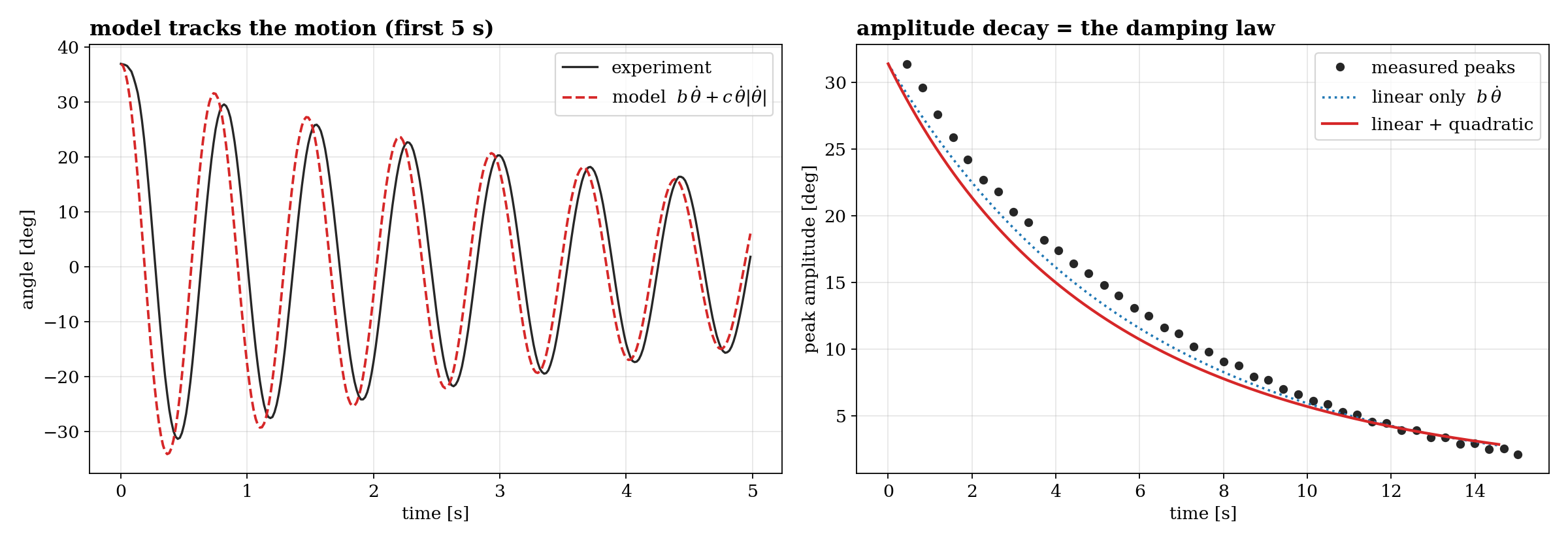}{Large-amplitude (${\sim}37^\circ$) free-decay
envelope: linear-only fit vs.\ the joint linear+quadratic fit that resolves $b$ and $c$
self-consistently.}

\subsection{Parameter identification with ODIL on a synthetic three-pendulum cradle}
\label{sec:odil-results}
All results in this subsection use the reduced model~\eqref{eq:reduced} with $N=3$, the release
protocol of the experiments (pendulum 1 held at $-0.6$~rad $=-34^\circ$ and released from rest,
pendulums 2--3 at the static repulsion offset), a horizon $T=20\sqrt{l/g}$ ($\approx$ 3 swing
periods) and $N_t=2000$ time steps ($h=0.01$). All computations run on a laptop CPU.

\begin{figure}[H]\centering
\includegraphics[width=0.78\linewidth]{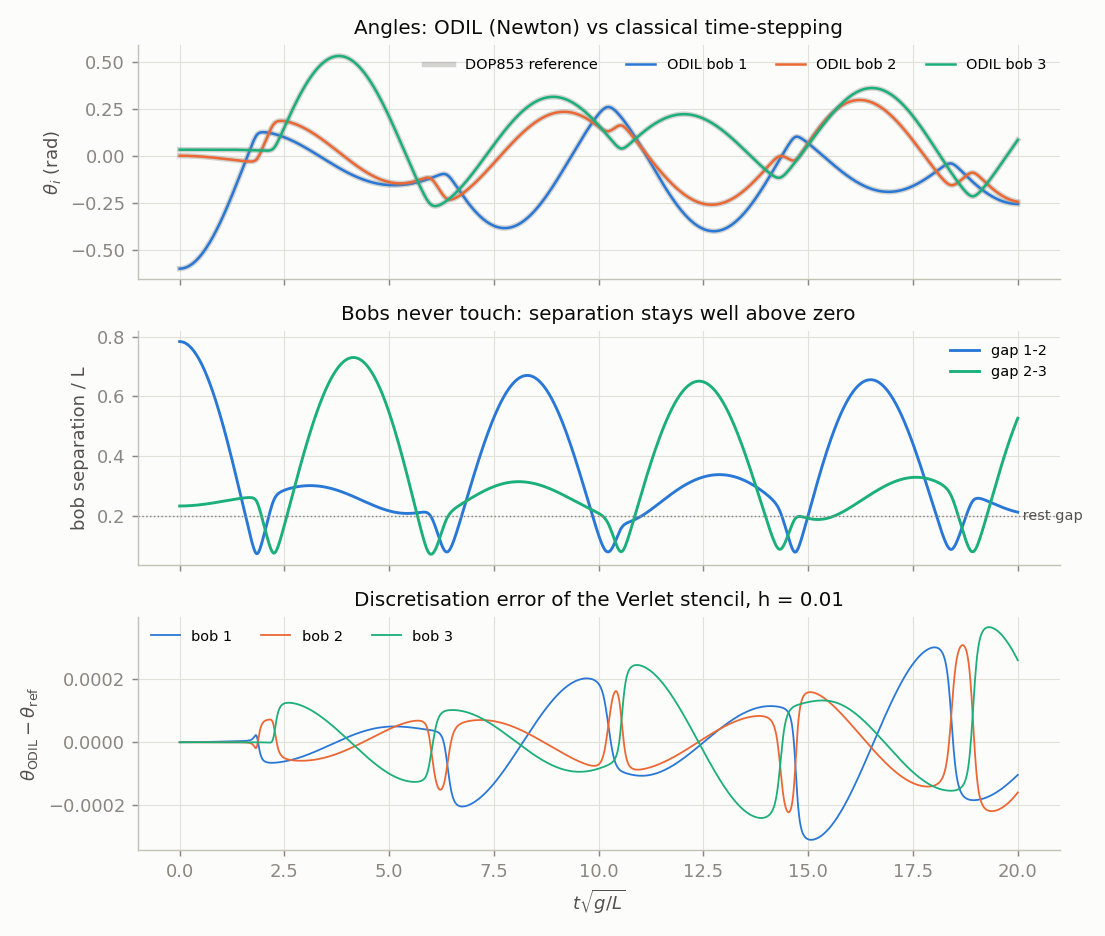}
\caption{Forward \ac{ODIL} solve of the three-pendulum cradle (point-dipole model). Top: angles from
\ac{ODIL} (color) over the DOP853 reference (gray). Middle: neighbor separations---the bobs never
touch. Bottom: \ac{ODIL} minus reference, i.e.\ the $O(h^{2})$ error of the Störmer--Verlet stencil at
$h=0.01$.}
\label{fig:odil-forward}
\end{figure}

\paragraph{Forward problem: ODIL reproduces the discrete dynamics to round-off.}
Starting from a constant trajectory and no time integration, the sparse Gauss--Newton solver
drives the discrete loss to $10^{-23}$ in 276 iterations over the three grids
($500\to1000\to2000$ steps; \SI{7}{s}), Fig.~\ref{fig:odil-forward}. The \ac{ODIL} trajectory agrees
with the explicitly time-stepped Verlet solution of the same stencil to $2\times10^{-11}$~rad,
confirming that the optimizer finds the exact minimizer, and with the DOP853 reference to
$3.6\times10^{-4}$~rad, the $O(h^{2})$ error of the stencil itself. The dynamics are those of a
cradle: the released pendulum's momentum passes through the middle magnet to pendulum 3, which
swings out to $0.53$~rad at $t=3.8$, while the neighbor separation never drops below $0.07\,l$
(Fig.~\ref{fig:odil-forward}, middle). The first-order baseline behaves very differently: 1000
Adam steps followed by 5000 L-BFGS iterations on the same loss (at $N_t=500$) stall at a loss of
$10^{-1}$ with an $O(1)$ trajectory error (Fig.~\ref{fig:odil-conv}). The stiff $1/r^{4}$
near-field force makes the problem badly conditioned for first-order methods, in line with the
observations of~\cite{karnakov2024odil}; all subsequent results use the Newton mode.

\fg[0.62\linewidth]{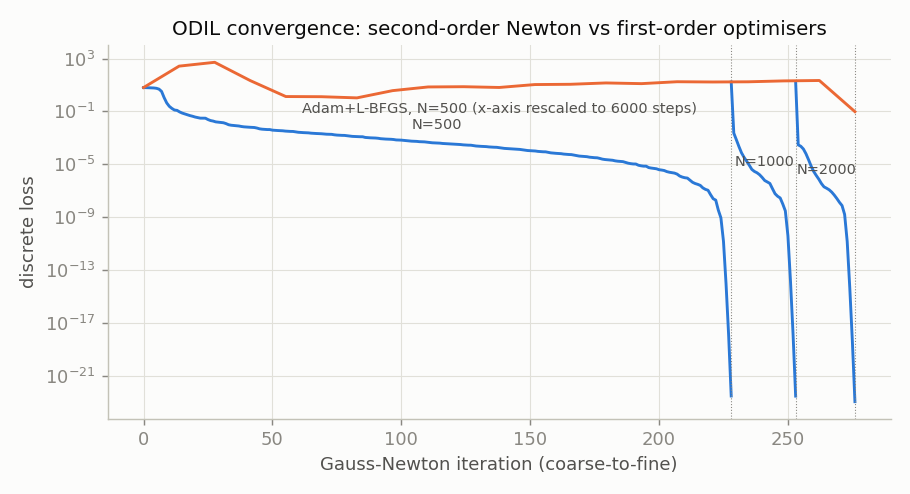}{Convergence of the discrete loss. Gauss--Newton
with the exact sparse Jacobian (blue) reaches round-off on each grid of the coarse-to-fine
hierarchy; Adam followed by L-BFGS (orange, abscissa rescaled to its 6000 steps) stalls at
$10^{-1}$.\label{fig:odil-conv}}

\begin{figure}[H]\centering
\includegraphics[width=0.56\linewidth]{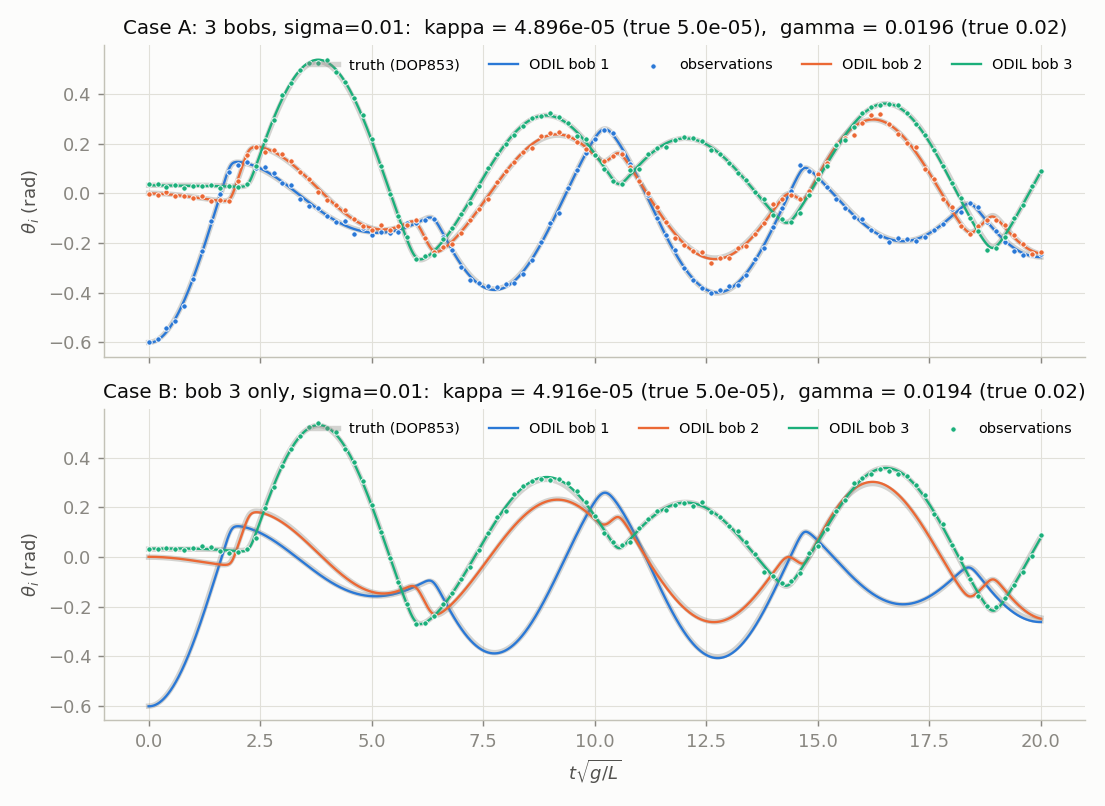}\hfill
\includegraphics[width=0.43\linewidth]{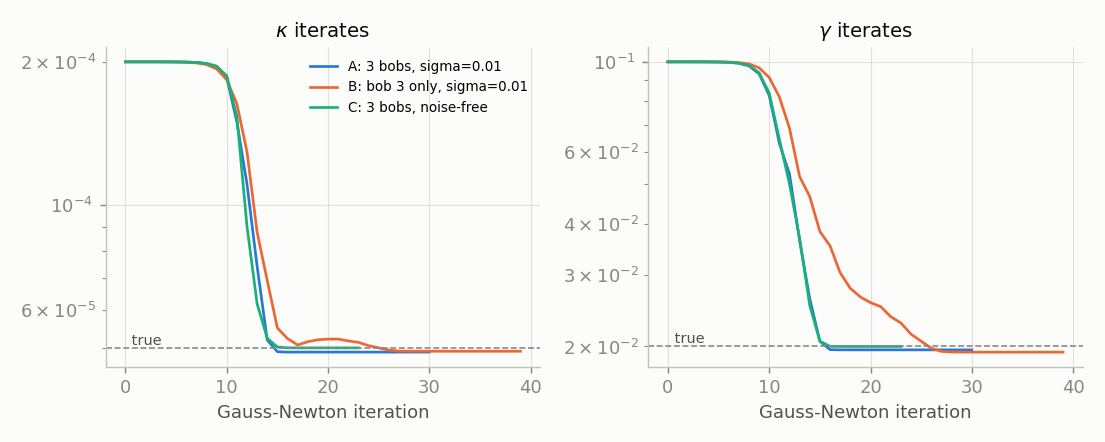}
\caption{Left: \ac{ODIL} trajectories (color) fitted to noisy observations (dots) over the truth
(gray) for cases A (top) and B (bottom, only pendulum 3 observed). Right: iterates of $\kappa$ and
$\gamma$ for cases A--C; the dashed line is the true value.}
\label{fig:odil-inverse}
\end{figure}

\paragraph{Inverse problem: magnet strength and damping from sparse noisy angles.}
Synthetic observations are taken every 20th step (100 samples per observed pendulum over the
horizon) with Gaussian noise $\sigma=0.01$~rad~$=0.6^\circ$, comparable to the Tracker
digitization error of the experiments. The initial guess is deliberately poor:
$\kappa$ four times and $\gamma$ five times their true values. Table~\ref{tab:odil-inverse} and
Fig.~\ref{fig:odil-inverse} summarize the recovery. With all three pendulums observed both
parameters are identified to $2\%$ in 30 Gauss--Newton iterations (\SI{4}{s}); with only the far
pendulum observed---the pendulum that never directly touches the released one---the coupling is
still identified to $1.7\%$, because the physics residual propagates the information through the
unobserved pendulums. With noise-free data the residual errors ($0.05\%$, $0.02\%$) are the
discretization bias of the stencil and shrink with $h^{2}$. The parameter iterates
(Fig.~\ref{fig:odil-inverse}, right) travel together with the trajectory: nothing moves for the
first $\sim$10 iterations while the trajectory is being pulled onto the data, then both parameters
converge in a further 10--15 iterations.

\begin{table}[H]\centering
\begin{tabular}{llcccc}
\toprule
case & observed & $\sigma$ [rad] & $\kappa$ error & $\gamma$ error & GN its.\\
\midrule
A & pendulums 1--3 & 0.01 & 2.1\% & 1.9\% & 30\\
B & pendulum 3 only & 0.01 & 1.7\% & 3.1\% & 39\\
C & pendulums 1--3 & 0 & 0.05\% & 0.02\% & 23\\
B$'$ & pendulum 3 only$^{\dagger}$ & 0.01 & \multicolumn{2}{c}{spurious minimum ($\kappa,\gamma\to0$)} & 200\\
\bottomrule
\end{tabular}
\caption{\ac{ODIL} identification of the magnetic coupling $\kappa\propto\Imag^{2}$ and the damping
$\gamma$ from synthetic angle data sampled every 20th step, starting from $\kappa$ 4$\times$ and
$\gamma$ 5$\times$ too large. Cases A-C initialize the trajectory with a time-stepped solution for
the guessed parameters; B$'$ ($^{\dagger}$) initializes it by interpolating the sparse observations}
\label{tab:odil-inverse}
\end{table}

The one failure (case B$'$) is instructive. If the trajectory is initialized by interpolating the
sparse data the
solver settles in a spurious minimum in which the coupling and the damping are driven to zero and
the observed pendulum is fitted as a free pendulum. Initializing the trajectory instead with the
time-stepped solution for the \emph{guessed} parameters removes the
problem in every case we tried. For the experimental application this costs one forward
integration and is the recommended protocol.

\paragraph{Model discrimination}
Figure~\ref{fig:odil-forcelaw} shows the static force between a repelling pair for the two magnet
models at equal moment. At the rest gap ($d_p=0.2\,l$) the loop force is $98$, $93$ and $76\%$ of
the dipole value for $a/l=0.0125$, $0.025$ and $0.05$; at the closest approach the cradle reaches
during a momentum transfer ($r\approx0.06$--$0.07\,l$) it drops to $82$--$86\%$, $51$--$59\%$ and
$18$-$24\%$, the same near-field discrepancy identified in dimensional form in
Fig.~\ref{fig:loopdipole}. Solving the forward problem with \ac{ODIL} for each radius, Fig.~\ref{fig:odil-fwdcompare}, the $a/l=0.0125$ magnet is indistinguishable from a
dipole ($|\Delta\theta_3|\lesssim0.01$~rad over $T$), the $0.025$ magnet drifts by
$0.05$--$0.08$~rad, and the $0.05$ magnet is a visibly
different cradle: the first transfer to pendulum 3 arrives later ($t=3.96$ instead of $3.80$) and
smaller ($0.502$ instead of $0.530$~rad), the bobs approach to $0.042\,l$ instead of $0.070\,l$,
and the trajectories separate by $0.2$--$0.3$~rad within four periods.

\fg[0.8\linewidth]{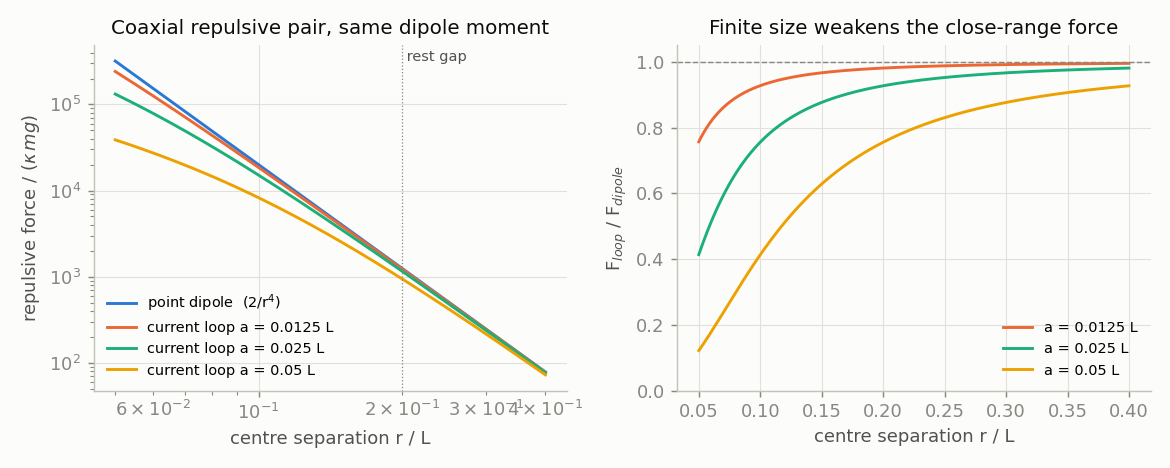}{Static repulsion between two magnets of equal
moment: point dipole ($2\kappa Mg\,(l/r)^{4}$) vs.\ exact current loops of three radii. Right:
force ratio; the finite size weakens the force by up to a factor of five at the separations the
cradle reaches during a momentum transfer.\label{fig:odil-forcelaw}}

\fg[0.8\linewidth]{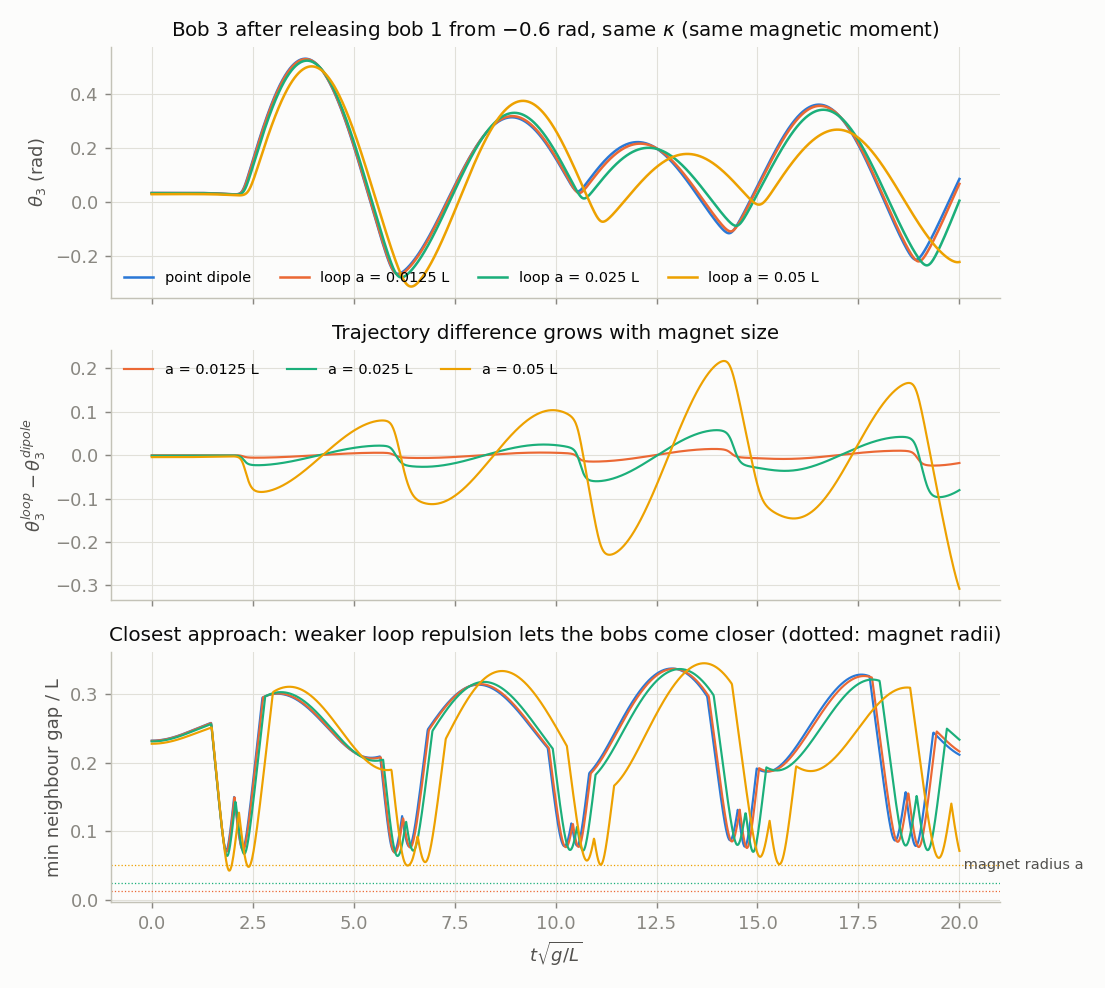}{Forward \ac{ODIL} solutions with identical $\kappa$ and
release for the point-dipole model and current loops of radius $a/l=0.0125$, $0.025$, $0.05$.
Top: pendulum 3. Middle: difference to the dipole solution. Bottom: closest neighbor separation
(dotted lines: magnet radii).\label{fig:odil-fwdcompare}}

The inverse problem turns this into a quantitative statement about model form error
(Table~\ref{tab:odil-discrepancy}, Fig.~\ref{fig:odil-discrepancy}). Fitted to noise-free data
produced by finite magnets, the point-dipole model can only lower its effective moment and raise
its damping: the identified $\kappa$ is $0.95$, $0.84$ and $0.64$ times the true value for the
three radii, i.e.\ an equivalent current $\Imag\propto\sqrt\kappa$ under-reported by $2$, $8$ and
$20\%$, and $\gamma$ is over-estimated by $1$, $6$ and $12\%$. What the effective moment cannot
absorb remains as an irreducible \ac{RMS} misfit of $4.5\times10^{-4}$, $1.9\times10^{-3}$ and
$7.8\times10^{-3}$~rad that is not noise-like but spikes at every momentum transfer
(Fig.~\ref{fig:odil-discrepancy}, right). For magnets of the size used here the dipole misfit is
therefore comparable to the digitization noise of a real measurement, so the dipole model would be
rejected by its residual and, if accepted regardless, would silently return a magnet current
$20\%$ low. The current-loop model fitted with $\kappa$, $\gamma$ \emph{and} $a$ free recovers all
three to better than $1\%$ from the same noise free data, from an initial guess with $a$ more than
three times too large in the smallest case, and its misfit ($\sim\!4\times10^{-6}$~rad) is at the level of the stencil
error. With $1\%$ measurement noise the loop model with $a$ known recovers $\kappa$ to $0.8\%$;
inferring $a$ as well gives $\kappa$ $5\%$ and $a$ $7\%$ low. The two parameters both control the
close-range force and are partly degenerate at this noise level, which is an argument for measuring
the magnet radius independently, as done in Table~\ref{tab:constants}, and for observing the
system where the transfers occur.

\begin{table}[H]\centering\small
\begin{tabular}{cccccc}
\toprule
& \multicolumn{2}{c}{dipole model fitted ($\kappa,\gamma$)} & \multicolumn{3}{c}{loop model fitted ($\kappa,\gamma,a$)}\\
\cmidrule(lr){2-3}\cmidrule(lr){4-6}
true $a/l$ & $\kappa_{\rm eff}/\kappa$ ($\Imag$ equiv.) & misfit [rad] & $\kappa,\gamma$ error & $a$ error & misfit [rad]\\
\midrule
0.0125 & 0.953 ($-2\%$) & $4.5\times10^{-4}$ & $<0.05\%$ & $+0.9\%$ & $4.6\times10^{-6}$\\
0.025  & 0.844 ($-8\%$) & $1.9\times10^{-3}$ & $<0.05\%$ & $+0.2\%$ & $3.7\times10^{-6}$\\
0.05   & 0.635 ($-20\%$) & $7.8\times10^{-3}$ & $<0.1\%$ & $+0.1\%$ & $3.2\times10^{-6}$\\
\bottomrule
\end{tabular}
\caption{Model discrimination with \ac{ODIL}. Noise-free data generated with the current-loop model are
fitted by both models from the same poor initial guess ($\kappa$ 4$\times$, $\gamma$ 5$\times$,
$a=0.04$). The dipole model absorbs part of the finite-size effect into a biased effective moment
and leaves the rest as an irreducible, structured misfit; the loop model recovers every parameter.
The rig's discs ($R/l=0.046$) correspond to the last row.}
\label{tab:odil-discrepancy}
\end{table}

\fg[0.98\linewidth]{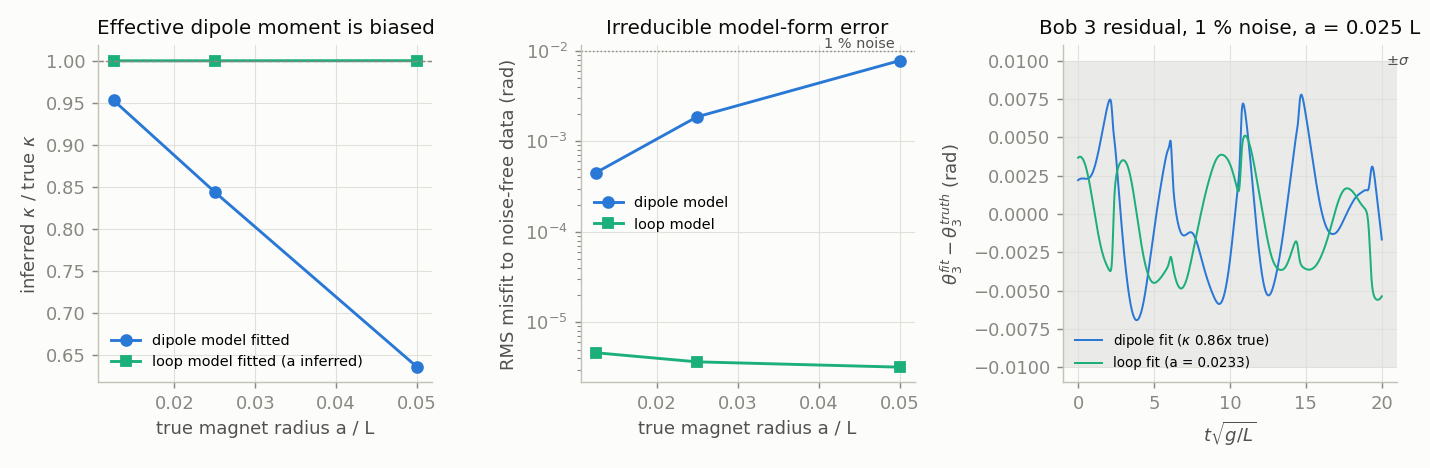}{Model-form error of the point-dipole approximation
quantified by \ac{ODIL}. Left: identified $\kappa$ relative to the true value vs.\ true magnet radius,
for the dipole (blue) and loop (green) models. Center: \ac{RMS} misfit to noise-free data; the dotted
line marks a $1\%$ ($0.6^\circ$) measurement noise. Right: residual of pendulum 3 for the two fits
to data with $1\%$ noise and $a/l=0.025$; the dipole residual (blue) spikes at every momentum
transfer, the loop residual (green) does not (shaded band: $\pm\sigma$).\label{fig:odil-discrepancy}}

\subsection{ODIL calibration on the measured trajectories}
\label{sec:odil-exp}
We now apply Algorithm~\ref{alg:odil} to the measured trajectories of the apparatus. The
primary data set consists of two independent recordings of the five-pendulum cradle at \SI{4}{cm}
pivot spacing (N45 discs of \SI{15.9}{mm} diameter and \SI{1.6}{mm} thickness pressed into the
printed enclosures with their axes along the row): in each, pendulum 1 was pulled out and released
and the angles of all five pendulums were tracked at \SI{59.94}{fps} for about \SI{31}{s}
(recordings DSC\,6824 and DSC\,6823, hereafter trials A and B). Trial A was exported from
Tracker at full precision; the angles of trial B were rebuilt from the tracked pixel positions
with the pivots of trial A, the camera not having moved between the two takes. Both records show
the static repulsion offsets ($(-5.3,\,-1.4,\,-0.3,\,+1.8,\,+5.1)^\circ$ in A and
$(-5.5,\,-1.2,\,-0.3,\,+1.6,\,+5.3)^\circ$ in B) and five clean spectral peaks at $1.44$, $1.83$,
$2.41$--$2.42$, $3.14$--$3.16$ and $3.95$--\SI{3.97}{Hz} whose cross-spectral shapes are the five normal
modes of \S\ref{sec:linmodes}. The older three-pendulum recordings at pivot spacings 2, 4, 6 and
\SI{8}{cm} (DP2--DP8), exported with a hand-placed origin per pendulum and angles rounded to
$0.1^\circ$, are analyzed with the identical protocol in \S\ref{sec:odil-compare}.

Every fit identifies, besides the physical parameters listed in the tables, the initial state, five
(three) rod frequency factors $f_k$ and five (three) angle-calibration offsets, the latter two with
Gaussian priors of $3\%$ and $1^\circ$. The quadratic damping coefficient is \emph{fixed} at its
single-pendulum value $c=\SI{2.7e-6}{}$ (\S\ref{sec:damping-results}) in every fit that does not list
$c$ among the identified parameters. The magnet strength and radius are identified on the first
\SI{10}{s} after release, where the signal-to-noise ratio is highest; the damping coefficients on the
first \SI{25}{s}. 

\subsubsection{Repeatability of the identified magnet}
\label{sec:odil-repeat}
\begin{figure}[H]\centering
\includegraphics[width=0.92\linewidth]{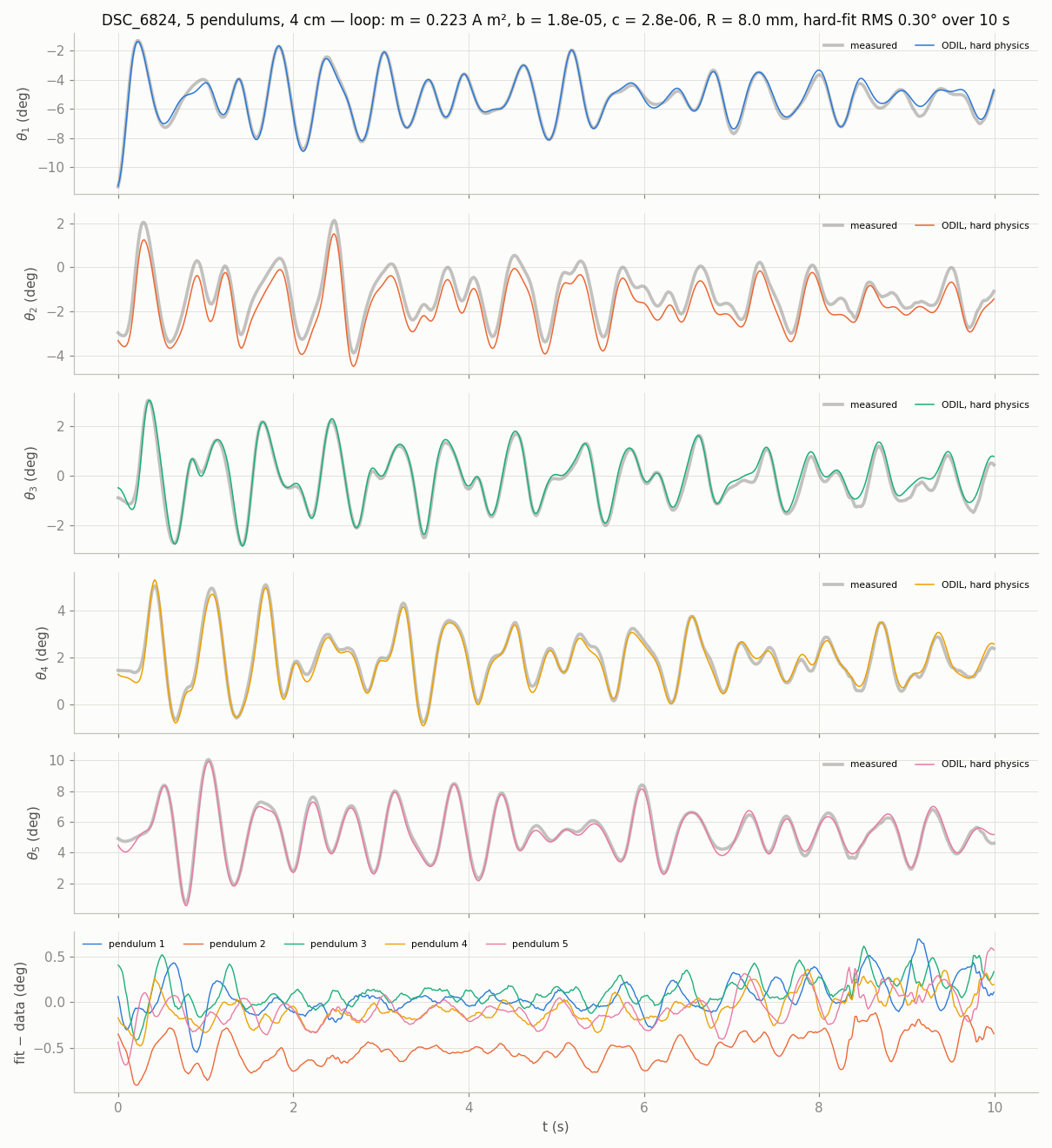}
\caption{\ac{ODIL} calibration of the five-pendulum cradle (trial A) with the current-loop model
(hard-physics stage): measured angles (gray) and the calibrated model trajectory (color) for the
five pendulums over the first \SI{10}{s} after release, and the residuals (bottom). The constant
offset of pendulum 2 is a calibration bias of the tracked angle; the dynamics of all five pendulums
are reproduced to $0.2^\circ$.\label{fig:odil-exp}}
\end{figure}

Table~\ref{tab:odil-exp} lists the parameters identified from the two five-pendulum trials side by
side. Three results follow. First, \textbf{the current-loop model identifies the magnet correctly and
repeatably}: it returns $\Imag=1122$ and \SI{1134}{A} from the two trials ($m=0.223$ and
\SI{0.225}{A.m^2}, a $1\%$ spread), within $3\%$ of the independent Hall-probe value of
\SI{1160\pm34}{A} and of the closed-loop dynamics fit of \S\ref{sec:validation}
(${\approx}\SI{1120}{A}$). The point-dipole model fitted to the same data needs $990$ and
\SI{1000}{A}, $15\%$ low in both trials, the same direction and similar magnitude as the bias found in the synthetic study of
\S\ref{sec:odil-results} for magnets of this size ($R/l=0.046$), although that study used a
point-mass model and larger releases. Second, \textbf{on full-precision data the magnet radius is
recovered from the motion alone}: with $R$ left free the two trials give $7.6$ and \SI{8.3}{mm},
bracketing the caliper value of \SI{7.95}{mm}. The  near-field force law of the real discs is
measurably different from a dipole's and the data promote  the loop model. Third, the calibrated loop model is \textbf{predictive}: it reproduces the outer
pendulums to $0.2$--$0.3^\circ$ \ac{RMS} over \SI{10}{s}, against an overall misfit of $0.35/0.41^\circ$ for the dipole model, and its linearized
modes ($1.817$, $2.404$, $3.121$, \SI{3.916}{Hz} for trial A) are within $1\%$ of the
measured peaks (Fig.~\ref{fig:odil-exp}). The one exception, pendulum 2 at $0.6^\circ$ in both
trials, is a constant $-0.5^\circ$ bias rather than a dynamical error: its measured rest angle sits
$0.4^\circ$ inside the mirror value of pendulum 4, a tracking-calibration offset that the $1^\circ$
prior did not allow the fit to absorb on the 10-s window (it is absorbed on the 25-s window, where
the data outweigh the prior, and the misfit of pendulum 2 drops to $0.4^\circ$).

\begin{table}[H]\centering\footnotesize
\begin{tabular}{lccccc}
\toprule
model & $m$ (A\,m$^2$) & $\Imag$ (A) & $R$ (mm) & $b$ ($10^{-5}$\,N\,m\,s) & \ac{RMS} misfit (deg)\\
\midrule
point dipole & 0.197 / 0.199 & 990 / 1000 & -- & 1.81 / 2.33 & 0.35 / 0.41\\
loop, $R$ measured & 0.223 / 0.225 & 1122 / 1134 & 7.95 & 1.85 / 2.35 & 0.30 / 0.35\\
loop, $R$ inferred & 0.220 / 0.227 & 1110 / 1145 & 7.6 / 8.3 & 1.84 / 2.36 & 0.30 / 0.35\\
\midrule
direct measurement & 0.230 & $1160\pm34$ & 7.95 & 2.65 & \\
\bottomrule
\end{tabular}
\caption{Repeatability of the \ac{ODIL} calibration on the two independent five-pendulum recordings
(\SI{4}{cm} spacing, first \SI{10}{s} after release, hard-physics stage); each cell gives trial A / trial B.
$\Imag=m/\pi R^{2}$ is quoted for $R=\SI{7.95}{mm}$ throughout so that the moments can be compared
directly with the Hall-probe value. The quadratic damping is fixed at its single-pendulum value
$c=\SI{2.7e-6}{}$. The initial state, five rod frequency factors and five calibration offsets were
identified alongside; the frequency factors stayed within $0.15\%$ of each other and the offsets
within $0.1^\circ$ of zero. \ac{RMS} misfit over all five pendulums; the per-pendulum values for the loop
model are $0.19/0.54/0.20/0.18/0.20^\circ$ (A) and $0.32/0.58/0.28/0.27/0.19^\circ$ (B).}
\label{tab:odil-exp}
\end{table}

\subsubsection{Identifying the damping}
\label{sec:odil-damping}
The direct measurement of \S\ref{sec:damping-results} resolved the linear and quadratic damping
coefficients from a single-pendulum decay at $37^\circ$. Can the cradle trajectories themselves
identify $c$? The discrete loss makes the test straightforward: $c$ (in the form $\gamma_2=c/I$) is
added to the list of unknowns and the 25-s window is fitted again. Table~\ref{tab:odil-damping}
collects the outcome for trial A with the loop model, together with the point-dipole result and
the corresponding fits for trial B.

\begin{table}[H]\centering\footnotesize
\begin{tabular}{llcccccc}
\toprule
trial & model & free & $b$ ($10^{-5}$) & $c$ ($10^{-6}$) & $b_f$ ($10^{-6}$) & \ac{RMS} misfit, pend.\ 1--5 (deg) & all\\
\midrule
A & loop & $b$ & 1.80 & 2.7 (fixed) & -- & 0.22 / 0.39 / 0.22 / 0.17 / 0.21 & 0.25\\
A & loop & $b,\,c$ & $\to0$ & 59 & -- & 0.24 / 0.40 / 0.22 / 0.18 / 0.23 & 0.26\\
A & loop & $b,\,b_f$ & 1.74 & 2.7 (fixed) & 1.6 & 0.22 / 0.39 / 0.22 / 0.16 / 0.21 & 0.25\\
A & dipole & $b,\,c$ & $\to0$ & 60 & -- & 0.39 / 0.33 / 0.24 / 0.18 / 0.30 & 0.29\\
B & loop & $b,\,c$ & $\to0$ & 70 & -- & 0.30 / 0.49 / 0.24 / 0.25 / 0.22 & 0.32\\
B & loop & $b,\,b_f$ & 2.25 & 2.7 (fixed) & 4.1 & 0.26 / 0.48 / 0.22 / 0.23 / 0.16 & 0.29\\
B & dipole & $b,\,c$ & $\to0$ & 71 & -- & 0.48 / 0.42 / 0.24 / 0.20 / 0.34 & 0.35\\
\midrule
direct (\S\ref{sec:damping-results}) & & & 2.65 & 2.72 & & & \\
\bottomrule
\end{tabular}
\caption{Damping identified on the 25-s window (hard-physics stage). ``free'' lists the damping
parameters identified together with $m$, the initial state, the frequency factors and the offsets;
$b$ in $10^{-5}$\,N\,m\,s, $c$ in $10^{-6}$\,N\,m\,s$^2$, $b_f$ (common-mode term, see text) in
$10^{-6}$\,N\,m\,s. ``$\to0$'' marks a parameter driven to the lower bound of its logarithmic
parametrization. The identified moment is $m=0.224$\,A\,m$^2$ in every loop fit of trial A.}
\label{tab:odil-damping}
\end{table}

With $c$ held at the single-pendulum value, $b=1.80\times10^{-5}$\,N\,m\,s reproduces the 25-s
record to $0.25^\circ$. Releasing $c$ does not improve the fit: the Levenberg--Marquardt iteration
wanders along a flat valley of the loss to the opposite corner: $b\to0$ and
$c\approx6\times10^{-5}$\,N\,m\,s$^2$, twenty to twenty-five times the single-pendulum value at an \ac{RMS} misfit of
$0.26^\circ$, indistinguishable from the linear fit. The same happens with the point-dipole model
and in trial B. A third form, a common-mode (``frame'') term $-b_f\,\overline{\dot\theta}$ acting on
the mean angular velocity of the row, which would model losses through the net reaction of the
pendulums on the support, is identified at $b_f=1.6$ and $4.1\times10^{-6}$ in trials A and B, a
tenth to a fifth of $b$, again with at most a marginal change of the misfit. The non-identifiability is expected from the amplitudes involved: the
trials were released from $6$--$7^\circ$, where the peak angular velocity is
$\dot\theta\approx1.2$--\SI{1.3}{rad/s} and the ratio of the quadratic to the linear torque at the
measured coefficients, $c\dot\theta/b\approx0.12$, has fallen to $0.03$--$0.04$ after \SI{10}{s}; the
single-pendulum decay of \S\ref{sec:damping-results} started from $37^\circ$, five times that
velocity, where the ratio is $0.6$, which is why it could resolve the two terms. The three-pendulum recordings
confirm the picture from the other side: at \SI{6}{cm} spacing the loop model returns
$c\to0$ with $b=3.0$--$3.3\times10^{-5}$, while the point dipole returns $b\to0$ with
$c=6$--$7\times10^{-5}$ on the same data (Table~\ref{tab:odil-wide})---the split between the two
terms is not a property of the cradle but of which force law is left to absorb the residual.

\begin{figure}[H]\centering
\includegraphics[width=0.98\linewidth]{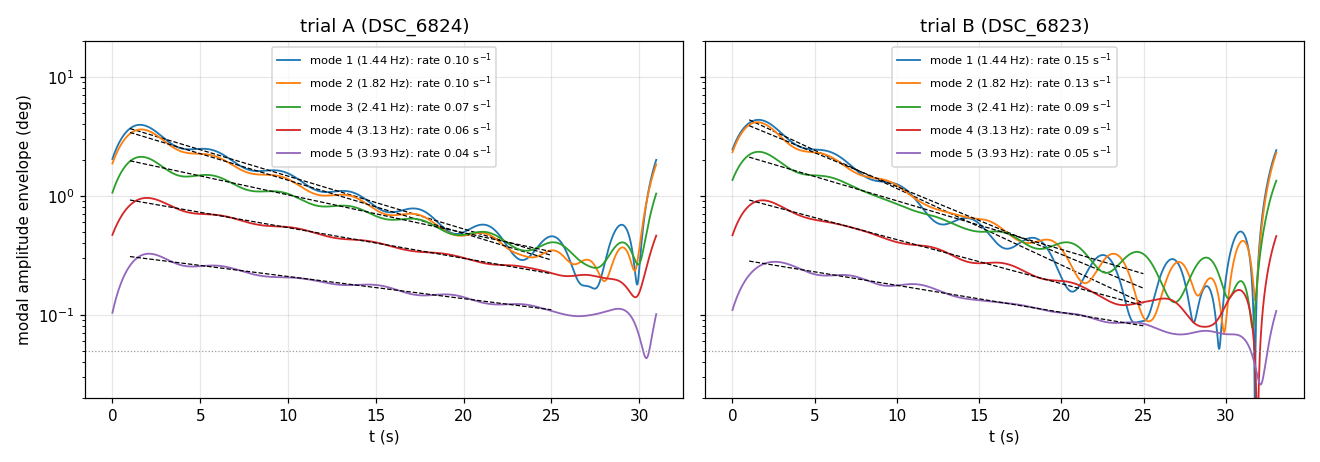}
\caption{Decay of the five normal modes in the two measured five-pendulum trials: envelopes of the
modal coordinates obtained by projecting the measured angles on the mode shapes of the calibrated
loop model, with exponential fits (dashed). Linear damping predicts the same rate for all modes;
the in-phase mode decays two to three times faster than the highest mode.\label{fig:odil-modal}}
\end{figure}

Projecting the measured angles on the five normal modes of the calibrated model and fitting an
exponential to each modal envelope (Fig.~\ref{fig:odil-modal}) gives decay rates of $0.10$ and
\SI{0.15}{s^{-1}} for the in-phase mode in trials A and B but only $0.04$-\SI{0.05}{s^{-1}} for
the highest mode, with the intermediate modes in between. A linear torque $-b\dot\theta$ damps every
mode at the same rate $b/2I$ (\SI{0.14}{s^{-1}} for the direct value); the observed rates
correspond to $b_{\rm eq}=1.9$--$2.7\times10^{-5}$ for the in-phase mode and
$0.8$--$1.0\times10^{-5}$ for the highest one. The quadratic term is the only parameter of the
present model that produces mode-dependent rates (the in-phase mode has the largest velocity
amplitude), which is the direction in which the free fit moved. However, the modal envelopes remain
exponential rather than hyperbolic, so the mechanism is more plausibly a loss that depends on the
mode shape---the in-phase mode is the one in which all five pendulums push on the support
together---than on the velocity squared. The misfit is insensitive to this because the fast modes
carry a few tenths of a degree of amplitude. A release in a pure high mode, or a cradle on a stiffer
support, would settle it; the identification machinery needs no change.

\subsubsection{Three- versus five-pendulum trials, and the directly measured parameters}
\label{sec:odil-compare}
The same protocol was applied to the five older three-pendulum recordings, whose pivot spacing
ranges from 2 to \SI{8}{cm}. Table~\ref{tab:odil-wide} sets the loop-model parameters identified
trial by trial against the directly measured values of Table~\ref{tab:constants} and
\S\ref{sec:damping-results}.

\begin{table}[H]\centering\footnotesize
\begin{tabular}{lcccccccc}
\toprule
 & direct & \multicolumn{5}{c}{three pendulums (spacing 2--8\,cm)} & \multicolumn{2}{c}{five pendulums (4\,cm)}\\
\cmidrule(lr){3-7}\cmidrule(lr){8-9}
 & & DP2 & DP4 & DP6(1) & DP6(2) & DP8 & A & B\\
\midrule
$\Imag$ (A), loop & $1160\pm34$ & 1376 & 1151 & 1120 & 1112 & 1269 & 1122 & 1134\\
$\Imag$ (A), dipole &  & 1288 & 1043 & 1050 & 1041 & 1219 & 990 & 1000\\
$m$ (A\,m$^2$), loop & 0.230 & 0.273 & 0.229 & 0.222 & 0.221 & 0.252 & 0.223 & 0.225\\
$R$ (mm), loop with $R$ free & 7.95 & 18.4 & 18.2 & 13.5 & 18.3 & -- & 7.6 & 8.3\\
$b$ ($10^{-5}$), $c$ fixed, 10\,s & 2.65 / 1.24 & 4.7 & 2.3 & 2.8 & 2.5 & 1.8 & 1.8 & 2.4\\
$b$ ($10^{-5}$), $c$ free, 25\,s & 2.65 & $\to0$ & $\to0$ & 3.3 & 3.0 & -- & $\to0$ & $\to0$\\
$c$ ($10^{-6}$), $b$ free, 25\,s & 2.72 & 249 & 90 & $\to0$ & $\to0$ & -- & 59 & 70\\
\ac{RMS} misfit (deg), loop, 10\,s &  & 0.85 & 1.23 & 0.34 & 0.45 & 0.67 & 0.30 & 0.35\\
static offset of pend.\ 1 (deg) &  & 9.3 & 5.0 & 1.9 & 2.2 & 1.3 & 5.3 & 5.5\\
\bottomrule
\end{tabular}
\caption{Parameters identified by \ac{ODIL}, trial by trial, against the direct measurements (hard-physics
stage; magnet parameters on the 10-s window, damping as indicated). DP2--DP8: three-pendulum
recordings at 2--\SI{8}{cm} pivot spacing, digitized with a hand-placed origin per pendulum and
angles rounded to $0.1^\circ$; A, B: the two full-precision five-pendulum recordings at \SI{4}{cm}.
The two direct values of $b$ are the large- and small-amplitude single-pendulum measurements. DP8
has the weakest coupling (static offsets $1.3^\circ$) and its free-radius and 25-s fits are
degenerate ($m\to0$), marked --.}
\label{tab:odil-wide}
\end{table}

Four observations follow. \emph{(i) The magnet strength is consistent across data sets.} The
three-pendulum trials at 4 and \SI{6}{cm} give $\Imag=1151$, $1120$ and \SI{1112}{A}, and the two
five-pendulum trials $1122$ and \SI{1134}{A}: five recordings, two apparatus configurations and two
digitization procedures agree to within $3.5\%$, all within $1.5\sigma$ of the Hall-probe mean
(\SI{1160\pm34}{A}), while the dipole values for the same recordings (990--\SI{1050}{A}) lie
3--5$\sigma$ below it. The two extreme spacings deviate from the Hall-probe value: DP2 (\SI{2}{cm}) gives $+19\%$ and DP8
(\SI{8}{cm}) $+9\%$. At \SI{2}{cm} the discs face each other across a gap of only $2.3R$ and the
identified moment is most sensitive to the geometry---the static offsets fix $m^2/d_p^4$, so
$m\propto d_p^2$ and a \SI{1}{mm} error in the pivot spacing or in the position of the magnet in its
enclosure moves $m$ by $10\%$; the \SI{2}{cm} record is also the one with the largest static
offsets ($9.3^\circ$), where the hand-placed tracking origin weighs most. At \SI{8}{cm} the coupling
is weak (offsets of $1.3^\circ$, with a $0.3^\circ$ asymmetry between the outer pendulums that the
model cannot produce), so the strength is identified from mode splittings of a few hundredths of a
hertz in angles rounded to $0.1^\circ$. \emph{(ii) The dipole bias is systematic.} The point-dipole
strength is $4$--$15\%$ below the loop value in every trial, so the $15\%$ bias of
\S\ref{sec:odil-repeat} is not a property of one recording. \emph{(iii) The magnet radius is the
parameter that exposes the data quality.} With $R$ free the full-precision five-pendulum exports
return $7.6$ and \SI{8.3}{mm}, the three-pendulum exports 13--\SI{18}{mm}. The radius is set by the
difference between the information in the static offsets and in the mode splitting; we attribute
the failure on the three-pendulum data to an angle-scale error of a few percent in the hand-digitized exports (\S\ref{sec:odil-repeat}) shifts
exactly that difference; the radius, unlike the strength, can only be read from full-precision
data. \emph{(iv) The linear damping agrees with the direct large-amplitude measurement.} With $c$
held at its single-pendulum value, $b$ is identified between $1.8$ and $2.8\times10^{-5}$\,N\,m\,s
in all trials at 4--\SI{8}{cm} (the \SI{2}{cm} trial, whose misfit is dominated by the near-field
model error, gives $4.7\times10^{-5}$), bracketing the direct large-amplitude value
$2.65\times10^{-5}$ and above the small-amplitude one ($1.24\times10^{-5}$). Between the two
five-pendulum trials $b$ differs by $25\%$ ($1.85$ against $2.35\times10^{-5}$); trial B was released
from a larger angle ($7.3^\circ$ against $6.0^\circ$ peak amplitude) and decays faster, the same
direction of amplitude dependence as the direct measurement.

\subsection{Prediction across 2–5 pendulums}
\label{sec:validation}
\begin{figure}[H]\centering
\includegraphics[width=0.5\linewidth]{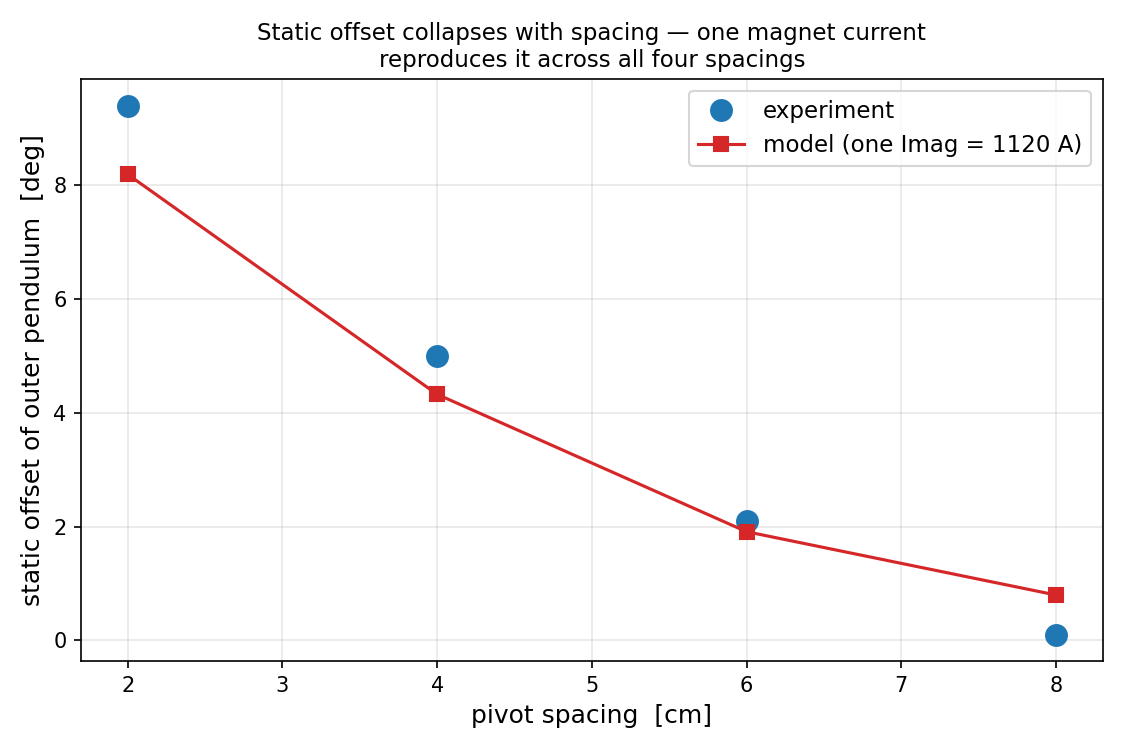}\hfill
\includegraphics[width=0.42\linewidth]{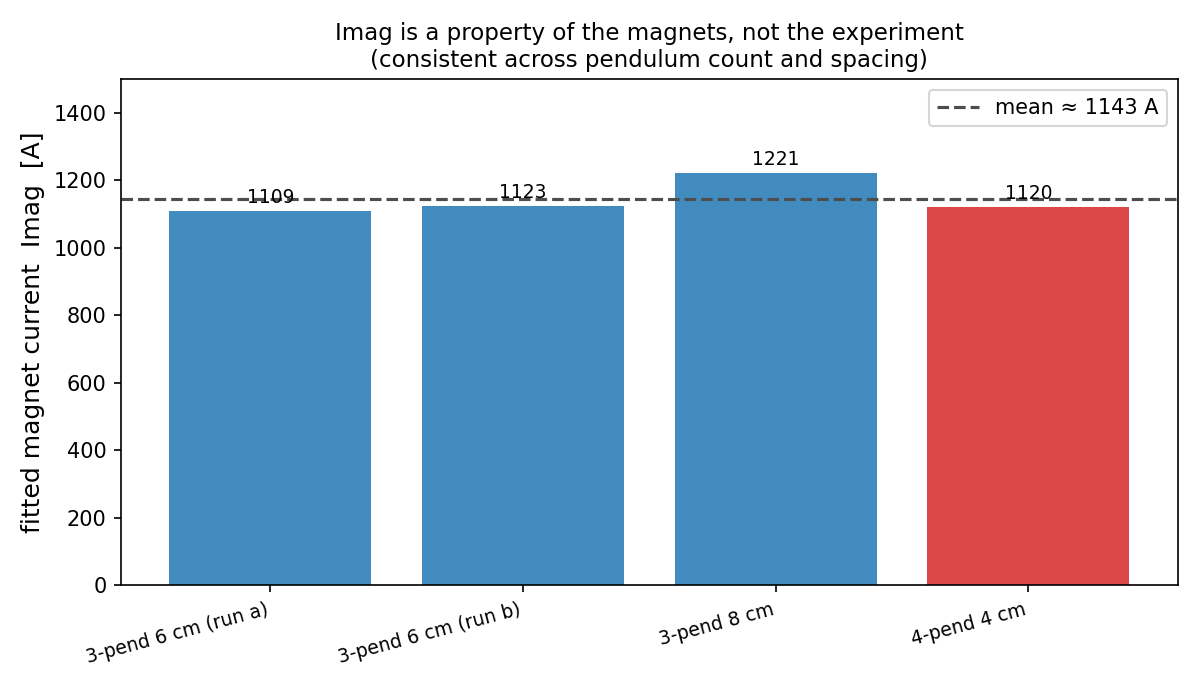}
\caption{Consistency checks. Left: the static repulsion offset of the end pendulum, produced by
the alternating-polarity mounting, collapses from $9.4^\circ$ at \SI{2}{cm} spacing to
${\approx}0^\circ$ by \SI{8}{cm}, reproduced with a single magnet current. Right: the fitted
$\Imag$ is essentially constant across pendulum count and spacing — a property of the magnet, not
of any one fit.}
\label{fig:consistency}
\end{figure}

With $R,\,l,\,\lcom,\,M,\,I$ measured, $b,c$ fixed from \S\ref{sec:damping-results}, and $\Imag$
fitted once to a three-pendulum recording and then transferred \emph{unchanged}, each configuration
is integrated closed-loop from its measured release angles alone — nothing magnetic is re-fit
between configurations, so reproducing a new pendulum count is a genuine prediction rather than an
independent fit. Fig.~\ref{fig:consistency} shows two consistency checks: the static repulsion
offset across pivot spacings and the fitted $\Imag$ across configurations.

\begin{table}[H]\centering
\begin{tabular}{lccl}
\toprule
pendulums & spacing & mean \ac{RMS} & notes\\
\midrule
2 & \SI{8}{cm} & $0.55^\circ$ & best of three 2-pendulum spacings tested (\SIrange{8}{10}{cm})\\
3 & 6--8\,cm   & ${\sim}0.50^\circ$ & two independent \SI{6}{cm} runs agree on $\Imag$ to ${\sim}1\%$\\
4 (single) & \SI{4}{cm} & $0.21^\circ$ & six interaction pairs\\
5 & \SI{4}{cm} & $0.28^\circ$ & ten interaction pairs\\
\bottomrule
\end{tabular}
\caption{Closed-loop prediction accuracy vs.\ pendulum count. One magnet, measured once, predicts
every configuration.}
\label{tab:validation}
\end{table}

\begin{figure}[H]\centering
\includegraphics[width=0.49\linewidth]{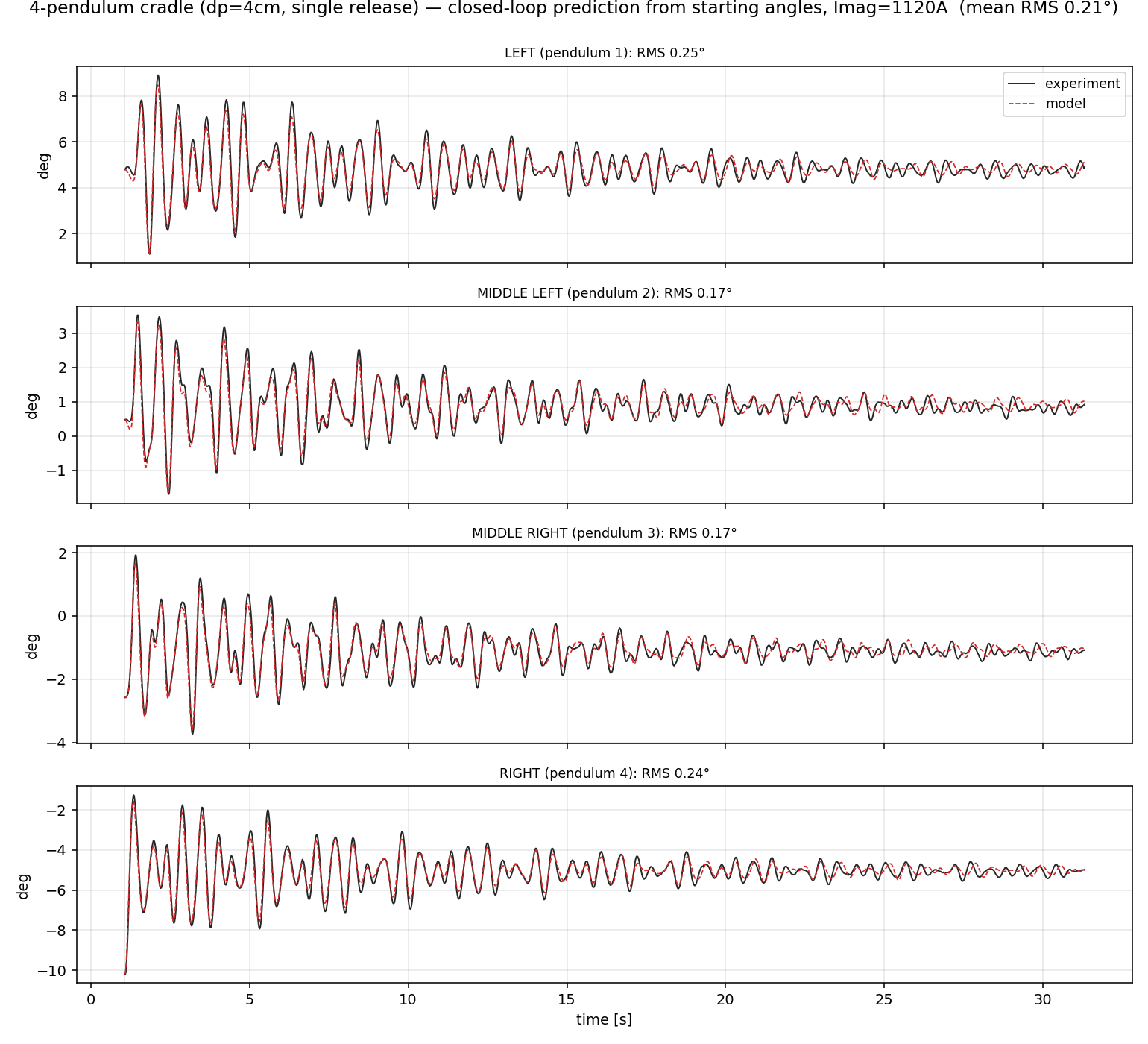}\hfill
\includegraphics[width=0.49\linewidth]{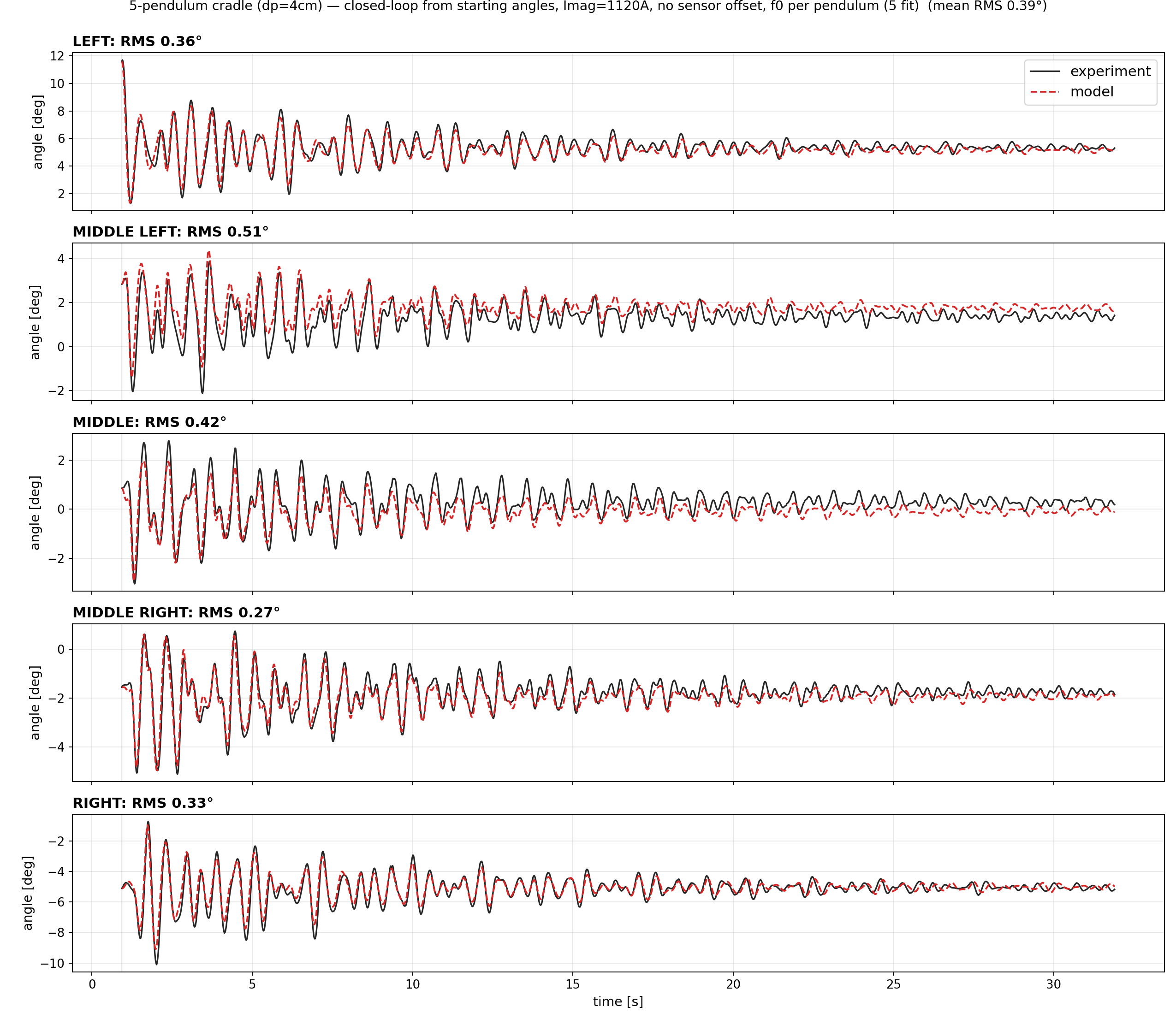}
\caption{Model (red, dashed) vs.\ measured (gray/black) angle traces. Left: 4-pendulum single
release, predicted to \SI{0.21}{\degree}. Right: five-pendulum, predicted to
\SI{0.28}{\degree}.}
\label{fig:results45}
\end{figure}

At large release amplitude the prediction error grows (Fig.~\ref{fig:limits}): larger swings bring
the magnets closest at their extremes, exactly where the finite-loop near-field force law is
least exact, and the motion becomes increasingly sensitive to the initial conditions.

\fg[0.58\linewidth]{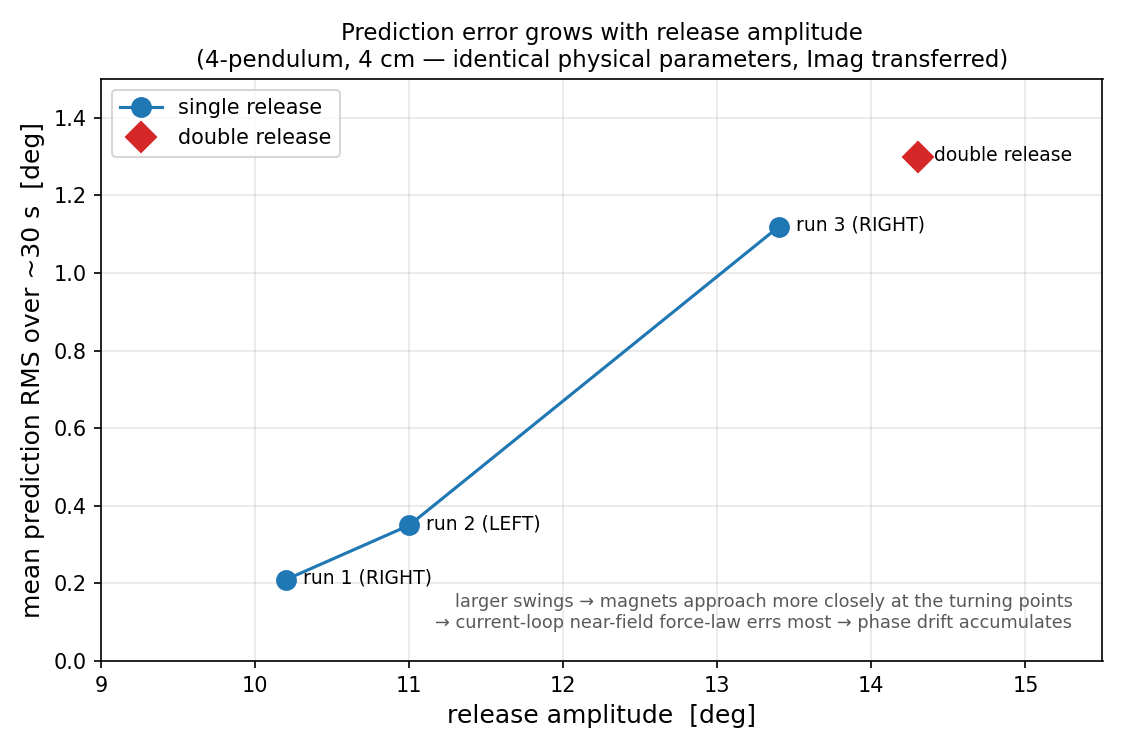}{\ac{RMS} prediction error vs.\ release amplitude.
The error grows as larger swings bring the magnets closer and make the motion more sensitive to
the initial conditions.\label{fig:limits}}

\subsection{Energy transfer in the MNC}
\label{sec:energy}
Like a classical cradle, energy released at one end of the chain travels to the other. The
mechanism differs: a classical cradle transfers energy \emph{discretely} — one ball in, one out,
the middle balls stationary — whereas the magnetic coupling is continuous, so energy spreads along
the chain as a beat, the same physical picture that underlies the normal-mode decomposition of
\S\ref{sec:linmodes}. Per-pendulum swing energy, $E_k=\tfrac12 I\dot\theta_k^2+Mg\lcom(1-\cos
\theta_k)$, is computed from the differentiated, one-period-smoothed tracked angle (removing the
within-swing kinetic$\leftrightarrow$potential exchange) and normalized to a fraction of the total.
This is a proxy for \emph{where} the energy is, not the full conserved budget: it omits the
magnetic interaction energy, which is shared between a pair of magnets and cannot be assigned to a
single pendulum, and which we separately recover by integrating the conservative torque,
\begin{equation}
U_{\rm mag}(\boldsymbol\theta)=-\Imag^{2}\sum_{\langle i,j\rangle}\!\int^{\boldsymbol\theta}\!\!
\big(\tau^{\mathrm{unit}}_{i}\,d\theta_i+\tau^{\mathrm{unit}}_{j}\,d\theta_j\big),
\end{equation}
giving $E_{\rm tot}=\sum_k\big[\tfrac12 I\dot\theta_k^2+Mg\lcom(1-\cos\theta_k)\big]+U_{\rm mag}$,
which is conserved up to the slow damping.

\fg[0.72\linewidth]{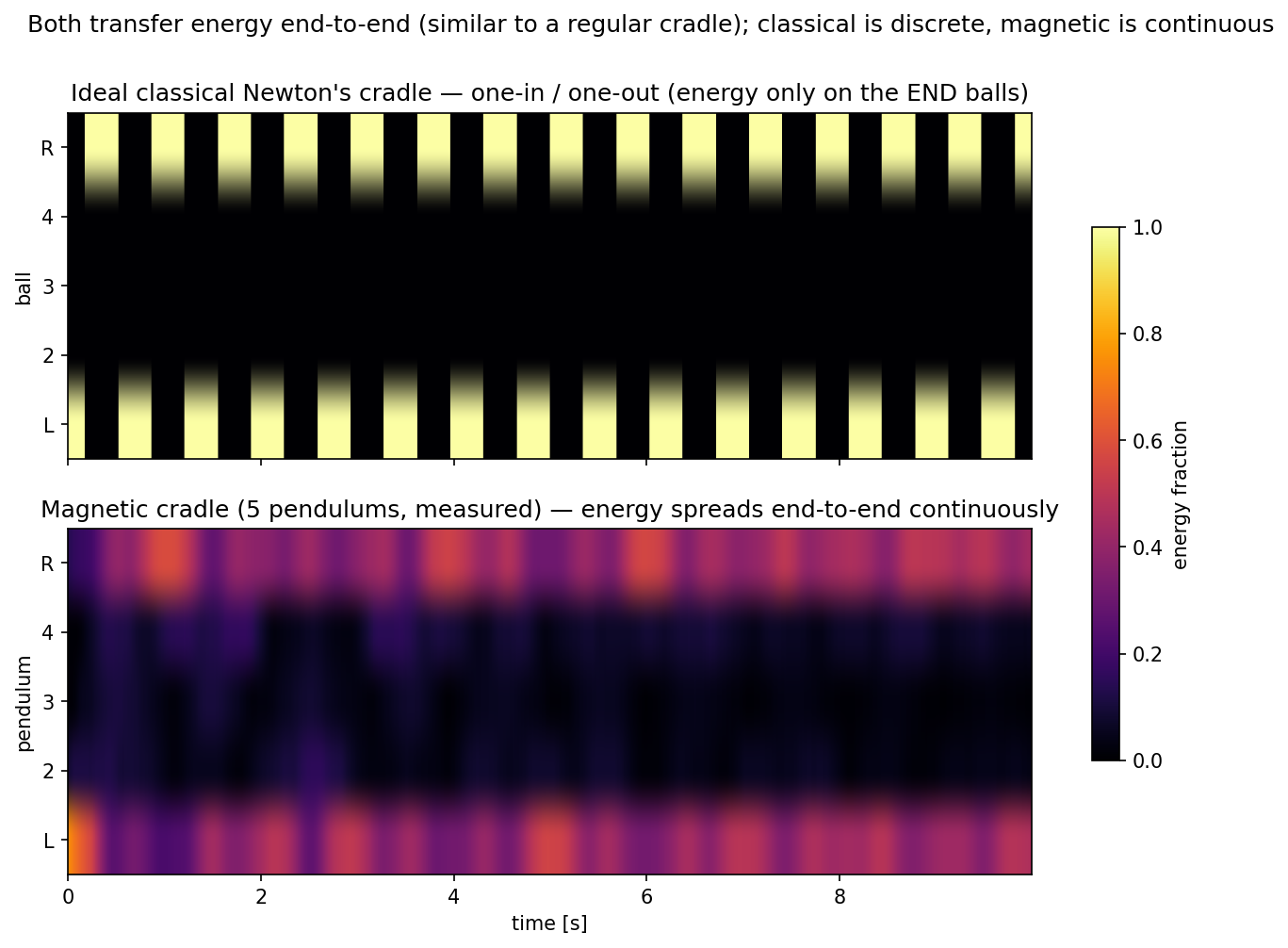}{End-to-end energy transfer. Top: an ideal
classical cradle (energy confined to the two end balls, simulated). Bottom: the magnetic
five-pendulum cradle (measured) — energy released at the left end spreads continuously down the
chain and back.}

\subsection{Linear normal-mode spectra}
\label{sec:modes-results}
The first-principles eigenfrequencies of~\eqref{eq:eig} — computed with no fitting to the Fourier
data — match the measured \ac{FFT} peaks to 1–2\% (Table~\ref{tab:modes}).

\begin{table}[H]\centering
\begin{tabular}{cll}
\toprule
$N$ & predicted modes [Hz] & measured \ac{FFT} peaks [Hz]\\
\midrule
3 & 1.45,\ 2.00,\ 2.87 & 1.46,\ 1.97,\ 2.81\\
4 & 1.45,\ 1.90,\ 2.59,\ 3.46 & 1.45,\ 1.88,\ 2.58,\ 3.44\\
5 & 1.45,\ 1.82,\ 2.41,\ 3.12,\ 3.92 & 1.45,\ 1.84,\ 2.41,\ 3.14,\ 3.95\\
\bottomrule
\end{tabular}
\caption{First-principles eigenfrequencies vs.\ measured \ac{FFT} peaks (\SI{4}{cm} spacing).}
\label{tab:modes}
\end{table}

\fg[0.58\linewidth]{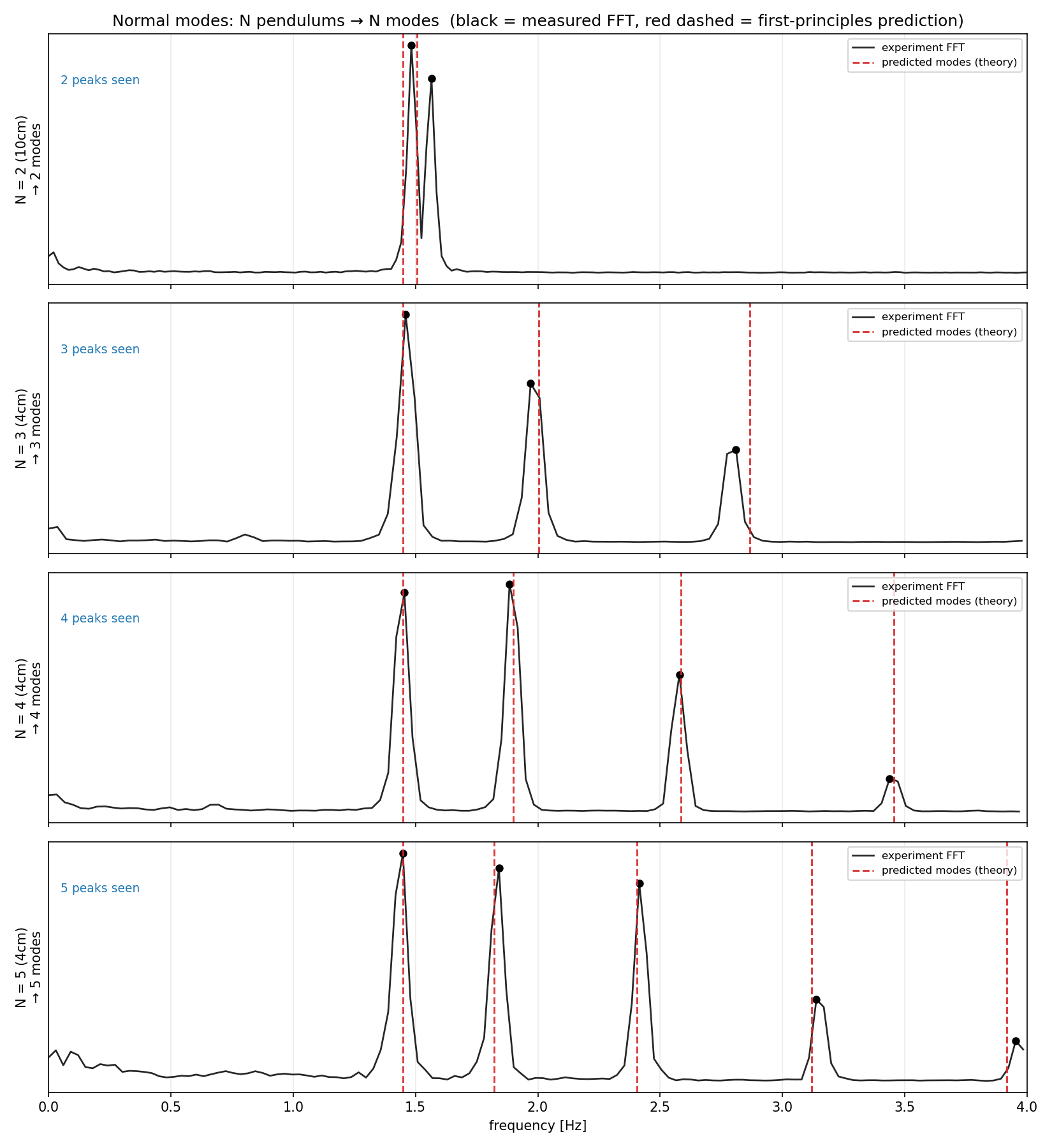}{Measured \ac{FFT} (black) with predicted mode frequencies
(red, dashed) for $N=2,3,4,5$. The lowest mode is the in-phase mode at ${\approx}f_0$; higher modes are magnetically stiffened and fan out as pendulums are added.}

The symmetry argument of \S\ref{sec:symmetry} is confirmed directly in the per-pendulum spectra
(Fig.~\ref{fig:participation}): the center pendulum of the 5-chain shows only its 3 symmetric
modes, with the antisymmetric peaks reduced to residual noise reflecting the small real-world
asymmetry (rod-to-rod variation of ${\sim}2\%$, imperfectly one-sided release) of an otherwise
mirror-symmetric rig.

\fg[0.55\linewidth]{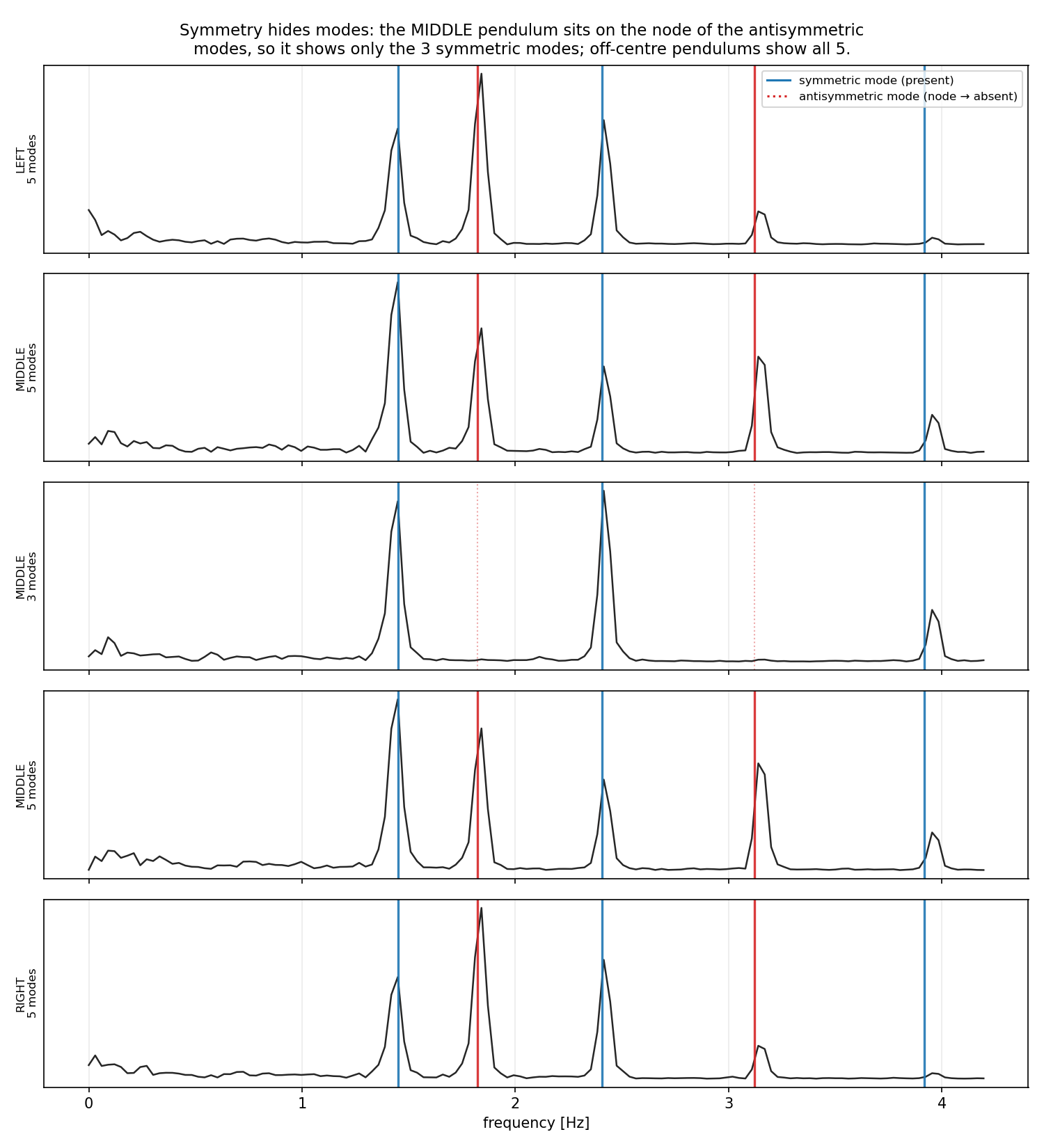}{Per-pendulum \ac{FFT} of the 5-cradle. Symmetric
modes (present for all pendulums) and antisymmetric modes (absent at the center, which sits on
their node) separate exactly as predicted by the reflection symmetry of
\S\ref{sec:symmetry}.\label{fig:participation}}

\subsection{Parameter study and the classical-cradle limit}
\label{sec:degeneracy}
Sweeping the model's eigenfrequencies against pivot spacing $d_p$, pendulum length $l$, and
magnet strength $\Imag$ (holding the rest of the model fixed) locates two universal facts.
First, the in-phase mode stays within $1\%$ of the free-pendulum frequency $f_0$ at every point
in parameter space, since it does not load the magnetic interaction. As the coupling vanishes
($d_p\to\infty$ or $\Imag\to0$) all $N$ modes collapse onto $f_0$: the zero-coupling limit is a
row of $N$ independent pendulums, with no energy transfer. The classical cradle lies at the
opposite extreme, an interaction of zero range and very high stiffness that lasts a short time
compared with the swing period; the magnetic interaction lasts a sizable fraction of a period, so
energy moves by beating between modes rather than by impulsive collisions. Second, coupling strength enters
symmetrically through spacing and magnet strength: decreasing $d_p$ or increasing $\Imag$ both fan
the higher modes upward from $f_0$, while increasing $d_p$ or decreasing $\Imag$ collapse them back
onto the degenerate uncoupled limit. Neither limit is singular: as $d_p\to0$ the model saturates to
a finite frequency (the finite loop size prevents divergence, and the magnets cannot physically
overlap below $d_p=2R$), and as $\Imag\to0$ every mode returns smoothly to $f_0$. Pendulum length
instead moves every frequency together as $l^{-1/2}$, since it rescales gravity rather than the
coupling.

\fg[0.6\linewidth]{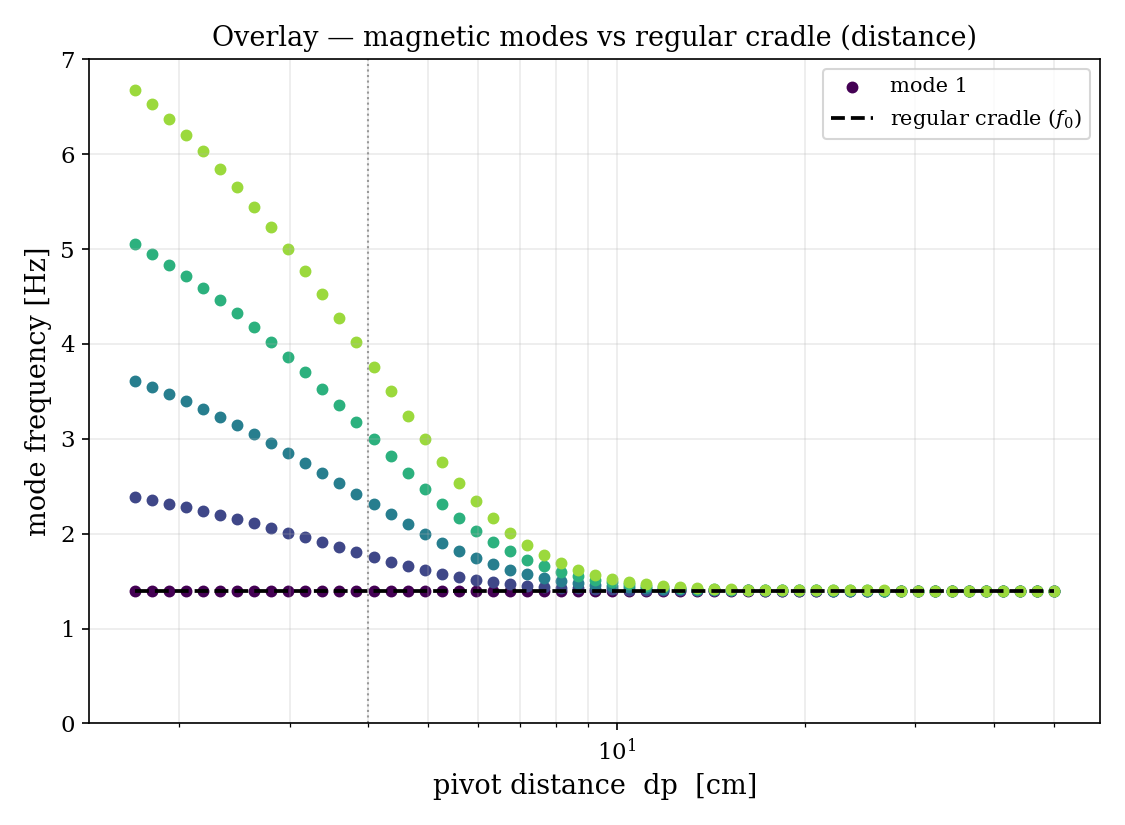}{Eigenfrequencies vs.\ pivot spacing:
magnetic cradle (5 modes, one color per mode), uncoupled pendulums ($f_0$, degenerate), and their
overlay. Coupling collapses onto the uncoupled limit as $d_p\to\infty$ and saturates (does not
diverge) as $d_p\to0$.}

\subsection{Limitations and Next steps}

The model is planar and treats the frame as rigid. Out-of-plane motion and frame compliance are
not represented; frame compliance is the most plausible origin of the mode-dependent damping of
\S\ref{sec:odil-damping}. The current-loop force law is itself the thin-disc
($L\ll R$) approximation to the true finite-height current sheet. For the \SI{1.6}{mm} discs used
here this is checked against, and agrees with, the exact finite-cylinder closed forms of~\cite{vokoun2009,ravaud2010}, but is not exact. The magnet strength $\Imag$ is now measured
directly by Hall probe (\S\ref{sec:hallprobe}), agreeing with the independent dynamics fit value
to within $4\%$. Per-pendulum
natural frequencies are calibrated per rig, since the rods of the three-pendulum rig differ by ${\sim}2\%$; an uncorrected error of this
size accumulates into several radians of phase over \SI{30}{s}. Finally, the classical cradle used for qualitative energy-transfer
comparison (\S\ref{sec:energy}) is an idealized, simulated one-in/one-out cradle and we have not  conducted experiments. This can be the topic of another investigation. Finally, the recordings analyzed here start at 6--$7^\circ$, where forward-shooting fits also
converge; demonstrating the advantage of \ac{ODIL} on large-amplitude, chaotic recordings is left
for future work. 

From the \ac{ODIL} study the identified magnet strength depends on the magnet model: fitting a point dipole to a
cradle built from $R/l\approx0.05$ discs returns a current $4$-$15\%$ low in every recording, and it is
the current-loop value that agrees with the Hall probe. That value is repeatable to $1\%$ between the
two five-pendulum recordings and to $3.5\%$ across the five-pendulum and the 4- and 6-cm
three-pendulum data sets, so the three determinations of $\Imag$---magnetostatic
(\SI{1160\pm34}{A}), closed-loop dynamics fit (${\approx}\SI{1120}{A}$) and \ac{ODIL}
($1112$--\SI{1151}{A}) are consistent to within $4\%$. The outlying 2- and 8-cm trials
mark the limits of the near-field geometry ($m\propto d_p^{2}$) and of weak coupling, respectively.
The  data term of the
discrete loss is a sharp diagnostic of data quality: the same exports could not be made
consistent by any model, and the inconsistency disappeared with a full-precision export. The same machinery answers the question of which damping the cradle can identify. Released as an
unknown, the quadratic coefficient $c$ is not resolved by trajectories that start at $6$--$7^\circ$:
linear, quadratic and common-mode forms of dissipation fit the 25-s records equally well
($0.25^\circ$), the free fit drifting to $b\to0$, $c\approx20$--$25\,c_{\rm direct}$ without any gain.
The identifiable quantity is the total dissipation, and the linear coefficient obtained with $c$
fixed at its single-pendulum value ($1.8$--$2.8\times10^{-5}$\,N\,m\,s across recordings) agrees
with the direct large-amplitude measurement.

\section{Conclusion}
We present a concise study of data driven parameter identification in a magnetic Newton's cradle model. 
Modeling each magnet as a finite current loop rather than a far-field point dipole, and
measuring, rather than fitting, every constant that can be measured independently, provides  a
magnetic Newton's cradle that is not only a  qualitative demonstration but a well quantified dynamical system.
A single magnet current — independently confirmed by direct Hall-probe field mapping to agree with
its dynamics-fit value to within $4\%$ — predicts chains of 2 to 5 pendulums to sub-degree accuracy
from their release angles alone. The same near-field model, linearized about
equilibrium, predicts the full normal-mode spectrum to 1–2\% with no fitting to the Fourier data 
and, cast as the residual of a discrete loss, allows the magnet strength, the damping, and the
magnet radius to be identified jointly with the trajectory by \ac{ODIL} — to within $1$--$2\%$ from sparse
noisy synthetic angles, and on the measured recordings at 4 and \SI{6}{cm} repeatably to within $4\%$ of the
Hall-probe current, with the magnet radius recovered to within $5\%$ on the two full-precision
recordings, with the linear
damping matching its direct measurement and the quadratic term shown to be unidentifiable at the
amplitudes recorded. 
The magnetic cradle shares the in-phase mode of the classical cradle, the one motion that does
not load the interaction. The
magnetic cradle differs from it not in whether energy transfers end-to-end, but in how and in
the richer spectrum of normal modes that the continuous, near-field magnetic coupling makes
possible.

Future work will investigate how a longer chain (8–12 magnets) can produce solitary-wave and single-site-localization regimes reported by Molerón et al.~\cite{moleron2014}. Extending the present study will allow  for a  system whose near-field force law is validated in closed
form rather than assumed. 

We believe that \ac{ODIL} provides a simple and quantitative way of introducing inverse problems as a methodology for the study of complex dynamical systems that can be easily understood and experimented with.

\appendix
\section{Implementation details of the ODIL solver}
\label{app:odil}
This appendix collects the formulas behind \S\ref{sec:odil} for readers who want to reproduce
or extend the solver; the accompanying notebook implements them verbatim.

\paragraph{Dipole force in closed form.}
With $\hat{\mathbf s}_k=s_k\hat{\mathbf x}$ the magnet axes and $\mathbf r_{ij}=\mathbf r_i-\mathbf r_j$,
the force on magnet $i$ that follows from~\eqref{eq:Udip}, $\mathbf f_{ij}=-\nabla_{\!\mathbf r_i}\tilde U^{\rm D}_{ij}$, is
\begin{equation}
\mathbf f_{ij}=\frac{1}{r_{ij}^{5}}\Big[(\hat{\mathbf s}_i\!\cdot\!\mathbf r_{ij})\hat{\mathbf s}_j+(\hat{\mathbf s}_j\!\cdot\!\mathbf r_{ij})\hat{\mathbf s}_i
+(\hat{\mathbf s}_i\!\cdot\!\hat{\mathbf s}_j)\mathbf r_{ij}
-\frac{5(\hat{\mathbf s}_i\!\cdot\!\mathbf r_{ij})(\hat{\mathbf s}_j\!\cdot\!\mathbf r_{ij})}{r_{ij}^{2}}\mathbf r_{ij}\Big],
\label{eq:fdip}
\end{equation}
and the torque about the pivot is $q_k=\big(\sum_{j\neq k}\mathbf f_{kj}\big)\cdot\partial\mathbf r_k/\partial\theta_k
=f_{k,x}\cos\theta_k+f_{k,z}\sin\theta_k$. This closed form is used only as a check of the
automatic-differentiation torque (agreement to $10^{-13}$).

\paragraph{Loop geometry and quadrature.}
Loop $i$ lies in the plane perpendicular to $\hat{\mathbf x}$ and extends in the out-of-plane
direction $\hat{\mathbf y}$: $\boldsymbol\rho_i(\varphi)=\mathbf r_i+a(\cos\varphi\,\hat{\mathbf y}+\sin\varphi\,\hat{\mathbf z})$
with tangent $\hat{\mathbf t}_i(\varphi)=-\sin\varphi\,\hat{\mathbf y}+\cos\varphi\,\hat{\mathbf z}$,
so $\hat{\mathbf t}_i\cdot\hat{\mathbf t}_j=\cos(\varphi_1-\varphi_2)$ and
\begin{equation}
\big|\boldsymbol\rho_i(\varphi_1)-\boldsymbol\rho_j(\varphi_2)\big|^{2}=\Delta x_{ij}^{2}+a^{2}(\cos\varphi_1-\cos\varphi_2)^{2}
+\big(\Delta z_{ij}+a(\sin\varphi_1-\sin\varphi_2)\big)^{2}.
\end{equation}
The double integral~\eqref{eq:Uloop} is replaced by
\begin{equation}
\tilde U^{\rm L}_{ij}\approx-\frac{s_is_j}{3\pi^{2}a^{2}}\Big(\frac{2\pi}{Q}\Big)^{2}
\sum_{p=0}^{Q-1}\sum_{q=0}^{Q-1}\frac{\cos(\varphi_p-\varphi_q)}{|\boldsymbol\rho_i(\varphi_p)-\boldsymbol\rho_j(\varphi_q)|},
\qquad \varphi_p=\frac{2\pi p}{Q},\quad Q=24 .
\label{eq:Uloop-quad}
\end{equation}
For coaxial loops at axial distance $z$ the exact result is
$\mathcal M=\mu_0a\,[(2/k-k)K(k^{2})-(2/k)E(k^{2})]$ with $k^{2}=4a^{2}/(4a^{2}+z^{2})$;
the quadrature~\eqref{eq:Uloop-quad} matches it to a relative error of $10^{-12}$ for $a\le0.05$, $z\ge0.06$.
On the axis $\mathcal M\to\mu_0\pi a^{4}/2z^{3}$ as $a\to0$, which reproduces the dipole energy
exactly. A disc of finite thickness is represented by $n_L$ loops equally spaced over the
thickness, each carrying $1/n_L$ of the current.

\paragraph{Jacobian blocks.}
Differentiating~\eqref{eq:stencil} with respect to the three time levels it involves gives
$3\times3$ blocks
\begin{equation}
\frac{\partial\mathbf R^{n}}{\partial\boldsymbol\theta^{n\mp1}}=\Big(\frac1{h^{2}}\mp\frac{\gamma}{2h}\Big)\mathbf I,\qquad
\frac{\partial\mathbf R^{n}}{\partial\boldsymbol\theta^{n}}=-\frac{2}{h^{2}}\mathbf I+\operatorname{diag}(\cos\boldsymbol\theta^{n})
+\kappa\,\frac{\partial^{2}\tilde U}{\partial\boldsymbol\theta\,\partial\boldsymbol\theta}\Big|_{\boldsymbol\theta^{n}},
\label{eq:jac-blocks}
\end{equation}
so the Jacobian with respect to the trajectory is block-tridiagonal, and the only non-trivial
block is the Hessian of the magnetic energy (the magnetic part of the stiffness matrix in~\eqref{eq:eig}). The columns for the identified constants are
\begin{equation}
\frac{\partial\mathbf R^{n}}{\partial\log\kappa}=-\kappa\,\mathbf q(\boldsymbol\theta^{n}),\qquad
\frac{\partial\mathbf R^{n}}{\partial\log\gamma}=\gamma\,\frac{\boldsymbol\theta^{n+1}-\boldsymbol\theta^{n-1}}{2h},\qquad
\frac{\partial\mathbf R^{n}}{\partial\log a}=-\kappa\,a\,\frac{\partial\mathbf q}{\partial a}\quad(\text{Model L only}).
\label{eq:jac-par}
\end{equation}

\paragraph{Assembly by coloring.}
Rather than coding~\eqref{eq:jac-blocks}, the sparse Jacobian is obtained exactly from
forward-mode automatic differentiation. For each color $c\in\{0,1,2\}$ and each component
$k\in\{1,2,3\}$ a tangent vector $\mathbf V$ is formed with $V^{n}_{k'}=\delta_{kk'}$ at all time
levels $n\equiv c\pmod 3$ and zero elsewhere; one Jacobian--vector product $\mathbf J\mathbf V$
then returns, for every residual row, the derivative with respect to the single unknown of that
color and component that the row depends on. Nine products give all $\approx27N_t$ non-zeros
without cancellation; one further product per identified constant gives~\eqref{eq:jac-par}.
The result was verified against a dense automatic-differentiation Jacobian to $10^{-10}$.

\paragraph{Levenberg--Marquardt control.}
With $\mathbf g=\mathbf J^{\!\top}\mathbf r$ and a trial step $\delta$ from~\eqref{eq:gn}, the
gain ratio $\varrho=[\mathcal L(\mathbf z)-\mathcal L(\mathbf z+\delta)]/[-\delta^{\!\top}\mathbf g-\tfrac12\delta^{\!\top}\mathbf J^{\!\top}\mathbf J\delta]$
compares the actual to the predicted decrease. A step is accepted if $\mathcal L$ decreases;
$\lambda$ is then multiplied by $\max\{\tfrac13,\,1-(2\varrho-1)^{3}\}$, otherwise the step is
rejected and $\lambda$ multiplied by 4. Iterations stop when $\mathcal L<10^{-20}$ (forward) or
$10^{-16}$ (inverse), or when no step is accepted. The first-order baseline uses 1000 Adam steps
(learning rate $10^{-2}$) followed by up to 5000 L-BFGS iterations with a zoom line search on the
same loss.

\bibliographystyle{unsrt}
\bibliography{bibliography}

\end{document}